# An Arduino Giga-Based Multi-Frequency Electrical Impedance Tomography System for Electrolyte Media: Hardware Design, Statistical Voltage Analysis, and System Performance Evaluation

Zahra Karimi[1], Natalie Verdugo[1], Joseph Schwab[1], Hamid Ghaednia[1]

[1] Department of Orthopaedics and Sports Medicine, Cedars Sinai, Beverly Hills, CA, USA

Correspondence: hamid.ghaednia@csmc.edu and joseph.schwab@csmc.edu

## Abstract

Multi-frequency electrical impedance tomography (EIT) typically depends on dedicated signal-generation hardware and precision analogue front-ends, which constrains cost, size, and reproducible deployment. We present a 16-electrode multi-frequency EIT system in which an Arduino GIGA R1 WiFi , a dual-core STM32H747XI platform serves as the sole signal-generation and data-acquisition engine, supported only by a discrete current source built from operational amplifiers and passive components, three analogue multiplexers, and simple signal conditioning. The two Arduino cores partition timing-critical waveform generation from system control, electrode multiplexing, and voltage acquisition; a host PC is used only for configuration and offline processing. Silver-wire electrodes are arranged around a 3D-printed cylindrical tank, and complete electrode measurement frames are acquired at eleven frequencies between 1 and 180 kHz, with a median repeat-frame signal-to-noise ratio of 48 dB over the 10–180 kHz analysis band at approximately 0.18 Hz per single-frequency frame (≈60 s per 11-frequency sweep) and cross-session baseline drift of approximately 6%. The system was evaluated in potassium chloride phantoms containing either a solid shell ABS cylinder, giving a fixed strongly negative conductivity contrast, or regenerated-cellulose dialysis bags filled with KCl from 0.001 to 1.0 M, spanning positive and negative contrasts of varying magnitude. Boundary voltages were analyzed directly to recover inclusion position, contrast polarity, and contrast magnitude without solving the inverse problem. In a balanced cohort of 30 runs of a rigid insulating inclusion, ten at each of three positions across independent sessions, nearest-class-mean template correlation located the inclusion in 29 of 30 runs at 160 kHz under leave-one-session-out validation (87–97 % across 10–180 kHz; chance 33 %). In a balanced cohort of 42 dialysis-bag runs, six at each of seven concentration ratios, the polarity of the conductivity contrast was recovered in all 36 runs with genuine contrast, and the signed contrast projection was linear in the cylindrical contrast factor ($R^2$ = 0.68 across individual runs; $R^2$ = 0.89 for the aggregate fit shown in Figure 3) but saturated above roughly tenfold contrast, so that concentration is ordered rather than measured (Spearman ρ = 0.85). Position recovery for the bags followed contrast, from 6 of 6 at the largest ratios to 2 of 6 at the matched condition, where no contrast exists. The matched-conductivity dialysis bag remained detectable (RMS perturbation ≈ 1.7 % against a 0.4–0.6 % noise floor); these results indicate that commodity microcontroller hardware can support usable multi-frequency EIT, and that for selected detection and identification tasks direct boundary-voltage analysis may substitute for full reconstruction. This provides a controlled in vitro basis for future investigation of EIT for monitoring conductivity changes associated with muscle electrochemistry.

# 1. Introduction

## 1.1 Electrical Impedance Tomography: Principles and Context

Electrical Impedance Tomography (EIT) estimates the internal distribution of complex admittivity, comprising electrical conductivity and permittivity, within a bounded domain. This is achieved by solving an ill-posed inverse problem from transfer impedances measured at a surface electrode array, acquired in response to multiple alternating current patterns applied sequentially at the boundary (Adler and Holder, 2021). The approach is sensitive to any property that alters local conductivity, such as ion concentration, fluid content, or phase distribution, and has been investigated in a wide range of applications, including lung ventilation monitoring in intensive care (Frerichs et al., 2017; Shi et al., 2021), muscle monitoring and rehabilitation (Zheng et al., 2022), implant health monitoring (Ghaednia et al., 2020; Keiderling et al., 2023), industrial process tomography (Rymarczyk et al., 2019), multiphase flow characterization (Yao and Takei, 2017), and geophysical subsurface imaging (Loke et al., 2013).

In practice, small alternating currents are injected through surface electrodes, and the resulting voltages are measured at the remaining electrodes. Repeating this process across multiple injection patterns produces a dataset that encodes the boundary response of the domain. The spatial admittivity distribution is reconstructed by solving a finite-element–based inverse problem, which is inherently ill-posed and requires regularization to obtain stable solutions (Adler and Lionheart, 2006; Adler and Holder, 2021). As a result, the achievable spatial resolution is fundamentally limited, particularly for features located deep within the domain. Depending on the acquisition and reconstruction method, EIT may estimate absolute or differential conductivity and, when phase-resolved complex measurements are retained, reactive or permittivity-related components.

Many practical biomedical systems nevertheless use fixed-frequency measurements and report relative conductivity or impedance changes. EIT is particularly well suited to the study of electrolyte systems, as aqueous electrolytes are predominantly resistive over the frequency range typically used (1–200 kHz) (Grimnes and Martinsen, 2015; Grossi and Riccò, 2017), and their conductivity increases monotonically with ionic concentration at fixed temperature. Over a bounded concentration range, the relationship may be represented well by an empirical linear calibration, although strict proportionality is not expected because molar conductivity changes with concentration. This enables EIT and related impedance methods to probe spatial or temporal variations in electrolyte concentration within fluid-filled domains, with relevance to process monitoring, food-quality assessment, and selected biological-fluid or cellular sensing applications (Grossi and Riccò, 2017; Yao and Takei, 2017; Li et al., 2025).

### 1.2 Instrumentation Challenges and Open-Source Approaches

While the principles of EIT are broadly applicable across domains, their implementation in biomedical settings introduces additional constraints. A central challenge in biomedical EIT lies in instrumentation. Accurate measurements require a stable current source, frequency-agility in multi-frequency systems, together with low-noise voltage acquisition, precise synchronization between current injection and measurement, and, in multiplexed architectures, reliable electrode switching. In addition, electrode–tissue interface impedance can introduce significant measurement errors if not properly controlled or compensated. These challenges are further amplified in wearable and real-world biomedical deployments. Motion artefacts, time-varying electrode contact, and deformation of the underlying tissue introduce additional sources of variability (Olson et al., 2025; Pennati et al., 2023; Kim et al., 2024). At the same time, constraints on size, power consumption, and hardware integration limit the use of conventional precision analogue front-ends and shielding strategies. Maintaining measurement fidelity under these conditions requires careful co-design of hardware, electrode configuration, and signal processing. Addressing these requirements typically necessitates dedicated analogue front-ends, precision signal chains, and synchronous detection schemes. Research-grade platforms such as the Sciospec ISX-3 EIT (Sciospec Scientific Instruments GmbH, Bennewitz, Saxony, Germany) provide broad frequency agility and configurability, while clinical systems, including the Dräger PulmoVista 500 (Drägerwerk AG & Co. KGaA, Lübeck, Schleswig-Holstein, Germany) and Timpel Enlight 2100 (Timpel S.A., São Paulo, SP, Brazil) are optimized for real-time bedside ventilation monitoring. However, translating these capabilities to compact, wearable, or widely deployable biomedical systems remains a significant engineering challenge (Pennati et al., 2023).

Several published Electrical Impedance Tomography platforms have demonstrated that reliable EIT measurements can be achieved across a range of hardware approaches. These include fully custom acquisition systems such as the KHU Mark2 (Oh et al., 2011), open-source integrated systems such as OpenEIT (Rintoul, 2018) and the EIT-kit (Zhu et al., 2021), and hybrid architectures that combine custom circuitry with commercial instrumentation. For example, the UCL ScouseTom pairs a commercial Keithley 6221 current source (Keithley Instruments, Solon, OH, USA) and EEG amplifiers with custom electrode-switching hardware (Avery et al., 2017). Field Programmable Gate Array (FPGA)-based implementations have further extended these capabilities, most notably in the KHU Mark2.5, where frequency synthesis, demodulation, and multi-channel timing are implemented across dedicated FPGA modules (Wi et al., 2014). Despite this progress, many systems still depend on bespoke printed circuit boards (PCBs), expensive external instrumentation, or specialized firmware, and relatively few have been released with fully open-source hardware and software licenses. This limits both integration into compact or wearable form factors and reproducible deployment across laboratories. A meaningful gap therefore persists between high-performance research platforms and systems that are sufficiently low-cost, accessible and deployable for wider biomedical and translational use. In response to this gap, we explore a minimal-hardware EIT architecture in which a widely available microcontroller (Arduino GIGA R1 WiFi; Arduino S.r.l., Monza, MB, Italy) serves as the sole signal-generation and data-acquisition engine. To our knowledge, this platform has not been systematically evaluated in this role for multi-frequency EIT.

The dual-core architecture is particularly relevant to EIT and is explicitly leveraged in this work. The two cores are used to partition system functionality, with one core dedicated to the timing-critical task of waveform generation and the other managing system control, electrode multiplexing, and voltage acquisition. This separation reduces contention between concurrent processes that, on single-core systems, would otherwise share execution time and introduce timing constraints in precision impedance measurements.

### 1.3 Analysis of Boundary Measurements Without Image Reconstruction

Image reconstruction remains a conventional endpoint of EIT, in which an ill-posed inverse problem is solved to estimate the spatial distribution of conductivity or complex admittivity (Borcea, 2002). Although such maps support spatial interpretation and visualization of the domain's electrical properties, the reconstructed distribution, as noted in Section 1.1, depends on the choice of forward model and regularization strategy (Lionheart, 2004) and may be affected by limited resolution, modelling errors, and reconstruction artefacts (Brazey et al., 2022). Moreover, iterative nonlinear and three-dimensional reconstruction can be computationally demanding, requiring repeated forward solutions and Jacobian computations (Borsic et al., 2012). Yet some practical EIT applications may not require a full reconstructed map and may instead address more targeted questions, such as whether an inclusion is present or whether the overall electrical properties of the domain have changed. In clinical translation, for example, the objective may be to detect a pathological change, discriminate between conditions, or track progression toward an intervention threshold. Apnea and non-apnea events have, for example, been classified directly from boundary-voltage inputs without image reconstruction during classification, although reconstructed EIT images contributed to the clinical annotation of the data (Vahabi et al., 2021). This raises a fundamental question: what application-relevant information do the boundary voltages themselves hold, and how much of it can be extracted reliably without full image reconstruction?

Approaches that process EIT measurements without reconstructing a conductivity image are comparatively uncommon and largely application specific. The most direct strategy classifies raw or normalized boundary voltages without solving the inverse problem: support vector machine (SVM) detection of prescribed intracranial hemorrhages in numerical and saline-head-phantom measurement frames (McDermott et al., 2018), detection and classification of failure in piezoresistive bone cement from phantom voltage readings with optional principal component analysis (Keiderling et al., 2023), and apnea detection from the real, imaginary, and magnitude components of neonatal boundary voltages using ResNet50 feature extraction with an SVM classifier, with event labels assigned by experienced clinicians using reconstructed EIT patterns (Vahabi et al., 2021). Machine learning has also been applied after reconstruction, to conductivity images or image-derived features, as in classification of bladder filling state (Dunne et al., 2018) and frequency-dependent symmetry metrics for stroke (McDermott et al., 2020); this retains spatial interpretability but requires the ill-posed inverse solution to be completed first. Elsewhere, machine learning replaces the inverse solver itself, mapping boundary voltages to conductivity distributions from finite-element-generated training data (Rymarczyk et al., 2019; Aller et al., 2023), with transfer to physical hardware limited by the correspondence between simulated and experimental measurements (Biasi et al., 2022). Task-oriented formulations go further and predict only the quantity the application requires, such as inclusion position, reporting substantially lower energy consumption than iterative reconstruction (Ikuno and Kaneko, 2026).

### 1.4 Scope of This Work

This study has two primary objectives: (1) to develop and evaluate a simple, low-cost, multi-frequency electrical impedance tomography (EIT) system using commercially available components, and (2) to investigate the extent to which inclusion location and conductivity contrast can be characterized directly from boundary-voltage measurements without image reconstruction. To address these objectives, we conducted controlled in vitro experiments using electrolyte phantoms in a benchtop EIT tank, systematically examining the system's measurement performance and sensitivity to changes in inclusion position and conductivity contrast. Together, these efforts aim to establish an accessible EIT platform and demonstrate the potential of measurement-domain analysis as a computationally efficient approach for extracting spatial and conductivity-related information, providing a foundation for future compact and wearable EIT applications.

## 2. Results

### 2.1 System performance and signal quality

The Arduino Giga generated stable square-wave excitation and acquired complete 11-frequency EIT datasets throughout all experiments. Excitation frequency was verified by oscilloscope and matched the target at all eleven settings. A complete 11-frequency sweep (16 injection patterns × 16 electrodes = 256 measurements per frequency) was acquired and recorded in approximately 60 s. This corresponds to approximately 5.5 s per single-frequency 256-channel frame, an effective frame rate of approximately 0.18 Hz. At this rate, the present system is suitable for static targets; approximately one minute separates successive 11-frequency acquisitions.

Signal-to-noise ratio was estimated from N = 25 repeated empty-tank frames (three time-lapse series over two days, 0.01 M KCl), as detailed in Section 5.15 (Eqs. 1–2).

Using consecutive-frame differences reduces the contribution of slow drift. The median over the 224 usable channels was 48 dB across 10–180 kHz (46–50 dB by frequency, essentially flat) and fell to 38 dB at 1 kHz. Taking the standard deviation over the whole series instead, which retains drift, gives 42.5 dB at 10 kHz and 45.6 dB at 160 kHz: the first estimate mainly reflects short-interval repeatability, whereas the second includes longer-timescale drift (Supplementary Figures S1 and S2).

The ScouseTom and KHU Mark2.5 systems introduced in Section 1.2 report 77.5 dB and 80 dB respectively (Avery et al., 2017; Wi et al., 2014), higher than the present value. Those measurements were obtained with resistor phantoms, whereas the present value includes the silver-electrode/electrolyte interface; consequently, these SNR figures provide context rather than a like-for-like comparison.

The present voltage-acquisition protocol is intended for relative measurements rather than calibrated transfer-impedance estimation. Backgrounds differing 8.9-fold in conductivity (1391 against 12414 μS/cm) would, with constant injected current and otherwise identical conditions, produce a corresponding inverse change in voltage; the observed difference across 10–180 kHz is 15 % (mean |V| ratio 0.85), or $V \propto \sigma^{-0.07}$. Accordingly, gated, single-ended square-wave measurements are analyzed relative to session-specific references rather than interpreted as absolute conductivity estimates (Section 5.8).

A session is one continuous period at the instrument with a single tank fill, an undisturbed electrode array and at least one empty-tank reference frame. Empty frames differ by 0.4–0.6 % (RMS Δ at 160 kHz) between consecutive acquisitions but by 1.7–11.9 % (median 5.7 %) between days, against an inclusion signal of 2.7 %. For comparisons across sessions, baseline variation was more prominent than short-interval noise. We therefore reference every inclusion measurement to an empty-tank frame acquired in its own session (Eq. 3); the influence of pooled referencing is described in Section 3.6.

### 2.2 Statistical analysis of raw boundary voltages

The two inclusion types were used to address complementary questions. The solid shell ABS inclusion presents a single fixed, strongly insulating contrast. Because the ABS contrast does not vary, any difference between one ABS run and another can be considered as a difference of position, making the ABS cohort suitable for testing whether position can be recovered from boundary voltages and whether performance depends on excitation frequency. The dialysis bags provided deliberately varied conductivity contrasts, filled from 0.001 to 1.0 M against a fixed background, so that the inclusion is in some runs less and in others more conductive than the electrolyte surrounding it. This cohort was used to evaluate position recovery and the polarity and magnitude of conductivity contrast relative to the background.

The processing is similar for both position and conductivity recoveries and is summarized in Figure 1. A run consists of 256 measurements at each of eleven frequencies. The frame is referenced to an empty-tank frame from its own session to give the relative perturbation Δ (Eq. 3), and the 32 channels measured on an electrode that is itself carrying current are discarded, leaving 224. Everything reported in this section is derived from the 224-channel vector: its magnitude RMS Δ (Eq. 4) describes how large a disturbance the inclusion produces; its correlation with per-position reference templates answers where the inclusion is; and its signed projection β onto the ABS template for the same position (Eq. 5) gives the sign and size of the conductivity contrast. No further transformation, log scaling, standardization, or frequency-wise normalization is applied.

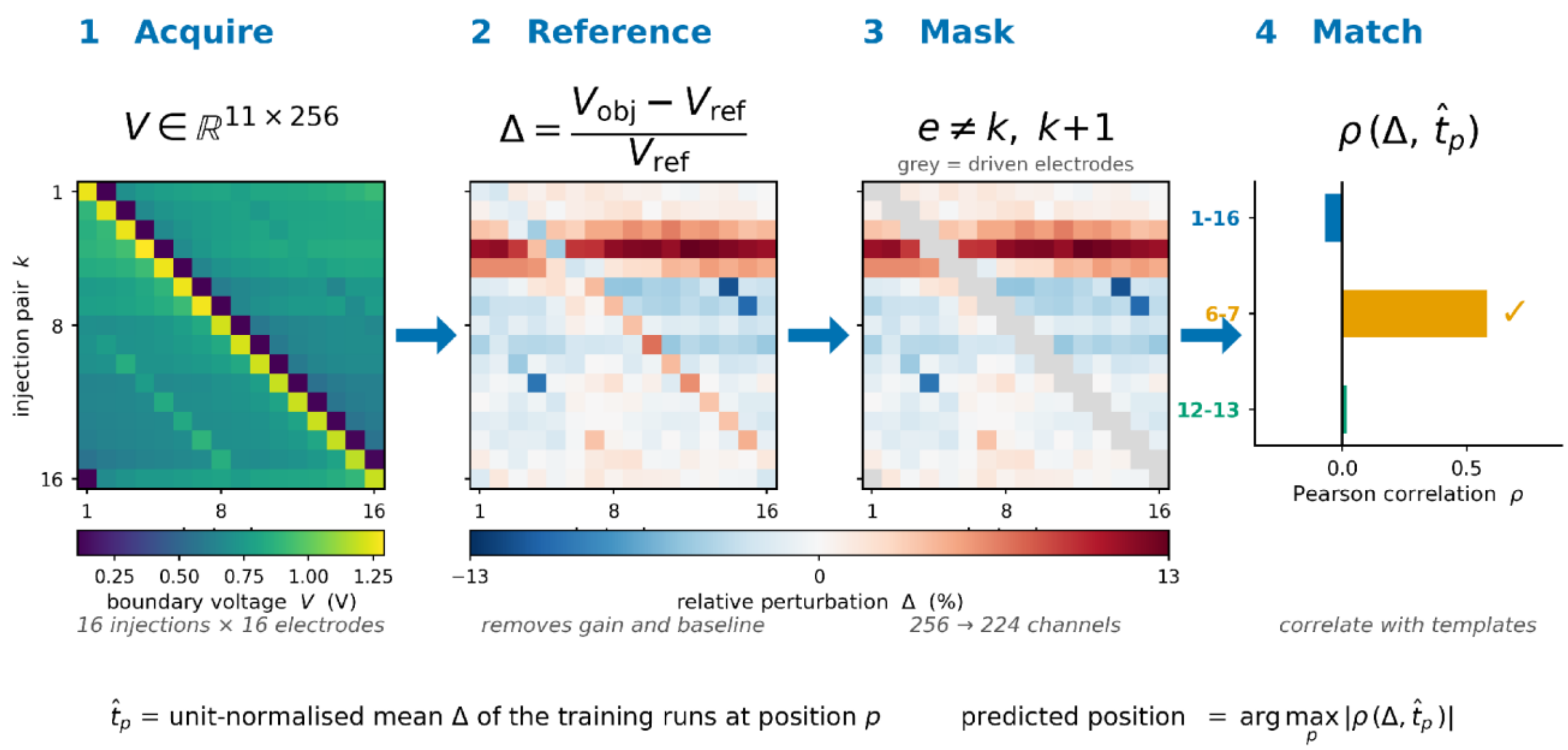

Figure 1. Boundary-voltage processing pipeline, worked through one measured run: the ABS inclusion at electrodes 6–7, 160 kHz, 0.01 M KCl background. (1) Acquire: 16 silver electrodes, adjacent-pair injection, giving 256 measurements at each of eleven frequencies; the panel shows the raw frame at 160 kHz as a 16 × 16 map of injection pair k against measurement electrode e, the bright diagonal being the driven electrodes. (2) Reference: division by the empty-tank frame of the same session gives the relative perturbation Δ (Eq. 3), removing channel gain and baseline; the inclusion appears as a coherent band on the injection pairs nearest it. (3) Mask: the 32 channels measured on an electrode that is itself carrying current are discarded (grey), leaving 224. (4) Match: Pearson correlation of Δ against the unit-normalized mean pattern $\hat{t}_p$ of each position, built from training runs only under leave-one-session-out; the largest absolute correlation identifies 6–7 correctly. The correlation shown, 0.58, is the median across the 30-run ABS cohort.

The primary analyses focus on 10–180 kHz based on signal quality and discrimination performance. Position classification at 1 kHz is 9 of 30, indistinguishable from the chance rate of 10 of 30, while the contrast relationship was weaker at this frequency ($R^2 = 0.30$ against 0.92–0.98 elsewhere, $F(6, 35) = 1.95$, $p = 0.10$). Perturbation magnitude at 1 kHz is simultaneously inflated by 2.7–11× relative to 160 kHz, a trend consistent with increased electrode-interface and baseline effects at lower frequencies; the present measurements do not isolate their individual contributions.

### 2.2.1 Recovering inclusion position

The rigid ABS inclusion was measured in 3 positions with 10 repetitions at each position (30 runs), spread across 10 sessions. Position was assigned under leave-one-session-out validation, in which every run from the test run's session was withheld from template construction. At 160 kHz the assignment was correct in 29 of 30 runs (96.7 %; 95 % CI 83–100 %; chance 33.3 %). Positions 1–16 and 12–13 were recovered in every run; the one failure was a 6–7 run assigned to 12–13 (Figure 2b).

Accuracy varied only weakly across the analysis band. Between 10 and 180 kHz it lay between 26 and 29 of 30 (87–97 %), a range of ten percentage points, with the maximum reached at five separate frequencies — 80, 100, 120, 160 and 180 kHz (Figure 2a). At 1 kHz accuracy fell to 9 of 30, below the chance value of 10 of 30. Perturbation magnitude behaved differently, and the contrast is instructive: the median RMS Δ declined gently across the band, from 3.3 % at 10 kHz to 2.6 % at 180 kHz, while accuracy did not fall with it, and at 1 kHz the perturbation was by far the largest recorded (14.8 %) at the frequency where accuracy was worst. What the classifier uses is therefore the shape of the perturbation pattern and not its size. Frequency selection within 10–180 kHz is consequently not critical for this task, showing comparable classification performance across the measured band under square-wave excitation.

Increasing the KCl background from 0.01 to 0.1 M raised the median inclusion-induced perturbation from 2.7 % to 4.5 % RMS Δ, while cross-background position classification remained more variable: templates built in 0.01 M identified 2 of 6 ABS runs acquired in 0.1 M at 160 kHz (chance: 2 of 6; 4 of 6 at 10 kHz). These exploratory observations from six runs across two sessions suggest that calibration and baseline conditions are important for cross-background transfer (Section 2.1).

Three further runs were acquired on day 2 with the inclusion placed at positions that were not among the three used for training and were analyzed without reference to their true location. Because the template classifier can only return one of the three trained classes, these runs were instead localized directly from the perturbation pattern, by taking the amplitude-weighted circular mean of the injection-pair index (Section 5.16). The estimate fell within one electrode pair of the true position in two of the three runs and within two pairs in the third, the poorest estimate belonging to the weakest perturbation (RMS Δ 1.7 % against 2.2–2.5 %). These exploratory measurements suggest that approximate angular position can also be estimated without a matching position template.

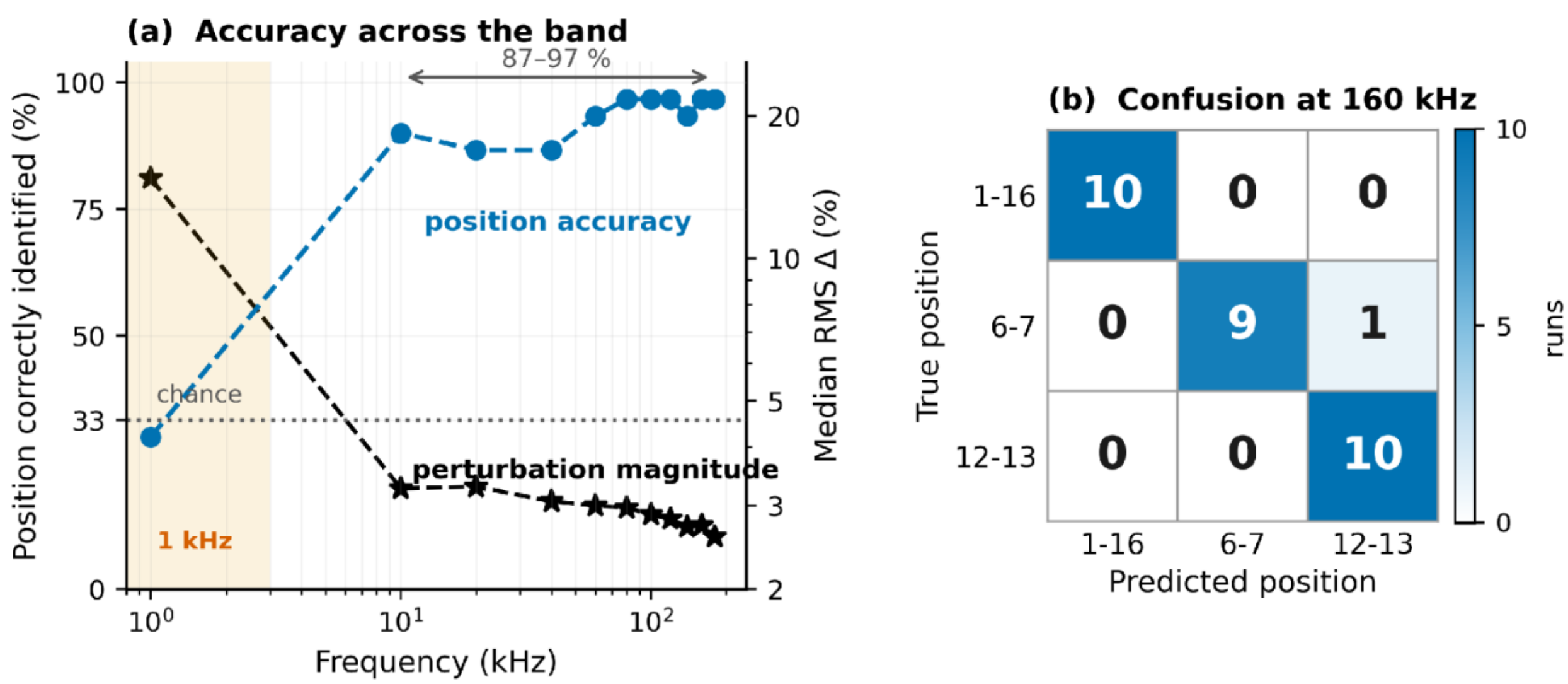

Figure 2. Recovery of inclusion position for the rigid ABS cylinder. (a) Percentage of runs whose position was correctly identified (blue circles, left axis) and median perturbation magnitude RMS Δ across the 30 runs (black stars, right axis, logarithmic), both plotted against excitation frequency, under leave-one-session-out validation. Accuracy lies between 87 and 97 % throughout 10–180 kHz while magnitude declines gently; at 1 kHz (orange shading, excluded from analysis) the perturbation is the largest recorded and the accuracy the lowest, indicating that what carries the positional information is the shape of the pattern rather than its amplitude. All dashed lines join measured points and are guides to the eye; the dotted line marks the chance rate for three classes (33.3 %). (b) Confusion matrix at 160 kHz. Rows are true positions, columns predicted positions; 29 of 30 runs are correct, the single error being one 6–7 run assigned to 12–13.

### 2.2.2 Recovering the conductivity contrast

The dialysis bags were filled from 0.001 to 1.0 M against a fixed 0.01 M background, giving inclusions both more resistive and more conductive than their surroundings. We first evaluated whether the sign of that contrast could be identified from the boundary voltages. In this cohort, it could: the signed contrast projection β was positive for every one of the 18 runs with a resistive bag and negative for every one of the 18 runs with a conductive bag, 36 of 36 in the balanced cohort. Polarity is the most robust quantity this instrument recovers.

Magnitude follows the contrast in an orderly way. β varied monotonically with inclusion conductivity, from +0.058 ± 0.008 (mean ± SEM) at 0.1× background to −0.182 ± 0.028 at 50× and −0.153 ± 0.035 at 100×, crossing zero close to the matched condition; one-way ANOVA across the seven levels gives $F(6, 35) = 14.18$, $p = 4.0 \times 10^{-8}$ (160 kHz, all three angular positions pooled, n = 42). Against the cylindrical contrast factor $C = (\sigma_{inc} - \sigma_{bkg})/(\sigma_{inc} + \sigma_{bkg})$, computed from measured conductivities, the relationship is linear: $\beta = -0.135 \times C - 0.031$, with $R^2 = 0.68$ across the 42 individual runs ($p = 1.4 \times 10^{-11}$) and $R^2 = 0.89$ for the aggregate fit annotated in Figure 3. These values describe the trend across concentration-level means and the variability of individual measurements, respectively (Figure 3).

The response appears to saturate above roughly tenfold contrast, but this is a property of the contrast factor and not of the instrument: C asymptotes to unity for a perfect conductor, reaching 0.95 at 50× and 0.97 at 100×, so there is little contrast left to resolve. The cylindrical form is the physically appropriate one because both inclusion types span the full electrolyte depth and the domain is quasi-two-dimensional; the spherical form $(\sigma_{inc} - \sigma_{bkg})/(\sigma_{inc} + 2\sigma_{bkg})$ was fitted for comparison ($R^2 = 0.970$ on level means, 0.716 on runs), while regression against log concentration ratio is worse ($R^2 = 0.696$ on runs).

Discrimination was strongest between broad contrast levels, with more overlap between adjacent concentrations. Of the 21 pairwise Tukey HSD comparisons among the seven levels, 11 are significant at $\alpha = 0.05$ and 10 are not, and none of the six adjacent pairs survives adjustment (0.1× vs 0.5×, p = 0.90; 0.5× vs 1×, p = 0.84; 1× vs 5×, p = 0.53; 5× vs 10×, p = 0.55; 10× vs 50×, p = 0.99; 50× vs 100×, p = 0.98); four non-adjacent pairs within the saturating range are also unresolved. Treated instead as a ranking task the projection orders concentration with Spearman ρ = 0.85 (n = 42). Together, these results support contrast-polarity identification and coarse magnitude ranking, while finer concentration discrimination remains a target for future refinement.

Two further observations bear on interpretation. First, the matched-conductivity bag still produced a measurable perturbation (median RMS Δ ≈ 1.7 %; range 0.7–4.3 %) against a 0.4–0.6 % noise floor, and the fitted intercept of −0.031 is consistent with that residual. This residual is consistent with a contribution from the membrane and displaced electrolyte; separating these effects will require additional controls (Section 3.4). Second, the contrast relationship is only weakly dependent on excitation frequency: the strength of the β–C relationship varies little from 10 to 180 kHz and is weaker at 1 kHz (Figure 4). Perturbation magnitude is likewise a weak function of frequency across that band, for the bags as for the ABS cylinder, so 160 kHz is a reporting choice rather than a necessary one.

Three features are evident in the channel-resolved patterns: the perturbation changes monotonically and in sign with inclusion conductivity, with resistive bags raising the boundary voltages in the injection groups adjacent to the inclusion and lowering them opposite while conductive bags mirror this; magnitude is largest in the injection groups flanking the inclusion and smallest remote from it (for the ABS inclusion, up to 5.4 % RMS Δ in flanking groups against ≈1.6 % remote, and for the 100× bag 2.8–3.0 % against ≈1.5 %); and the matched condition gives a pattern of much reduced but non-zero magnitude. Perturbation magnitude for the ABS inclusion was 2.7 % (RMS Δ, median across 30 runs at 160 kHz; interquartile range 2.1–3.8 %) with a median per-run maximum single-channel perturbation of 10.4 % (largest observed 30.5 %); the bags produced systematically smaller perturbations at every level, 0.9 % at 0.1× and 0.7 % at 0.5×, approaching the ABS value only at the largest contrast tested.

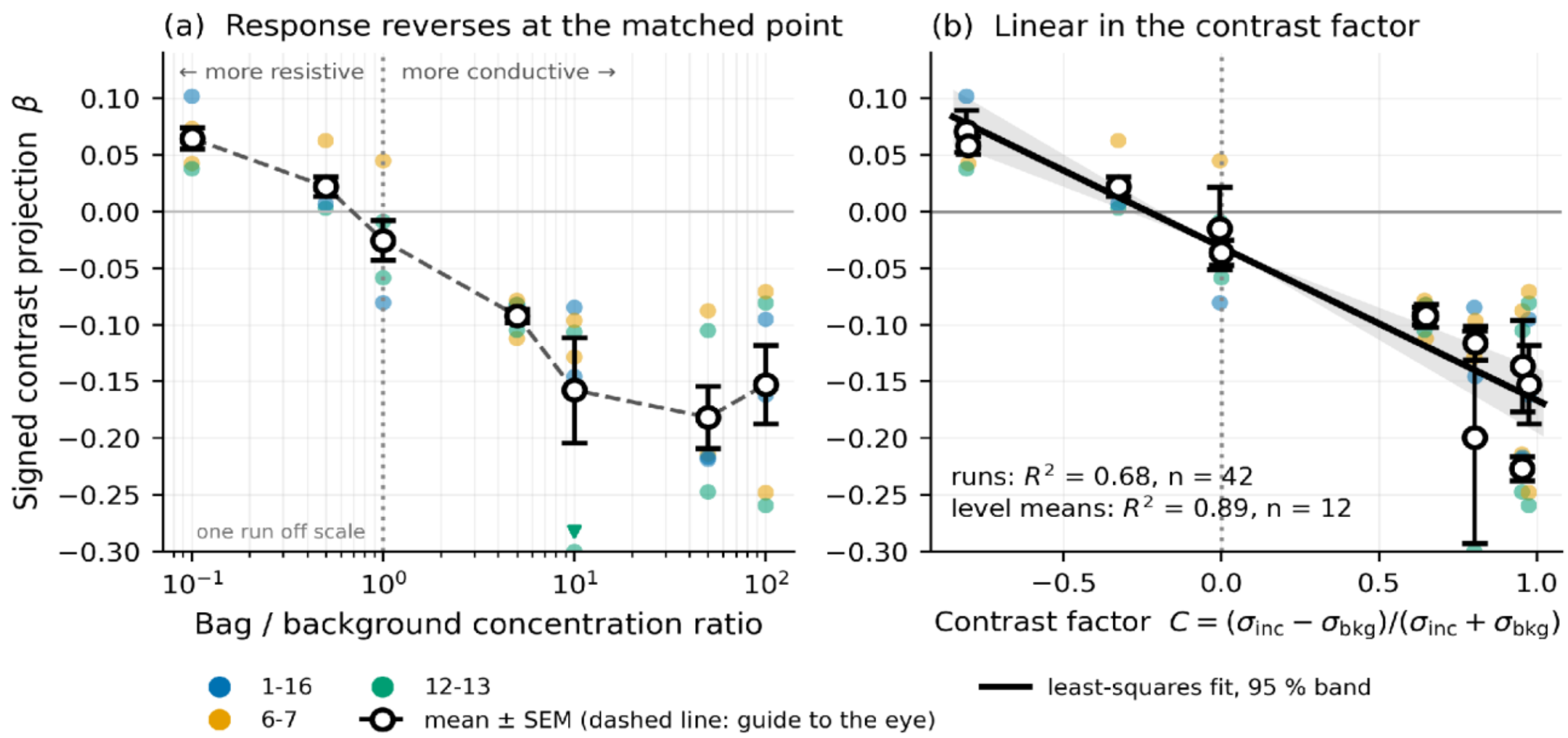

Figure 3. Signed contrast projection β at 160 kHz for the 42 dialysis-bag runs of the balanced cohort in 0.01 M background. (a) β against nominal concentration ratio; points are individual runs colored by position, open symbols with error bars are level means ± SEM, and the dashed line connects the means as a guide to the eye. The zero crossing falls at the matched condition. (b) β against the cylindrical contrast factor C computed from measured conductivities, with the least-squares fit and its 95 % band. The apparent saturation in (a) is the asymptote of C itself. One run at 10× lies below the plotted range and is marked by an arrow.

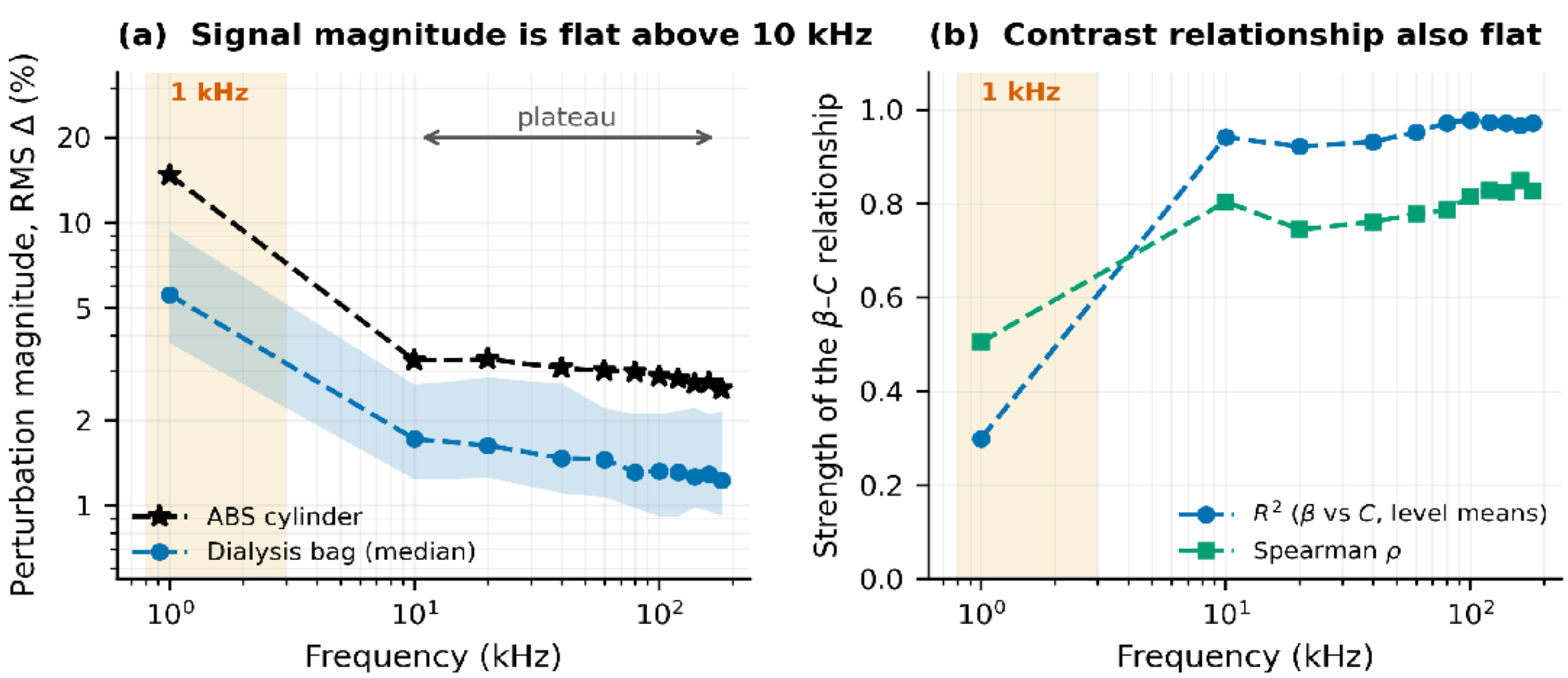

Figure 4. Perturbation magnitude and the contrast relationship are both only weakly dependent on excitation frequency above 10 kHz. Both panels pool the three angular positions and use the balanced cohorts (30 ABS runs, 42 dialysis-bag runs) in 0.01 M KCl. (a) Median perturbation magnitude against frequency for the ABS cylinder (black stars) and the dialysis bags (blue circles; shaded band is the interquartile range across the 42 runs). (b) Strength of the β–C relationship over the same band: $R^2$ on the seven concentration-level means (blue) and the Spearman rank correlation of β with concentration (green). Both vary little at the analyzed frequencies and collapse only at 1 kHz. Orange shading marks the 1 kHz point, outside the primary analysis band. All lines join measured points and are guides to the eye; none represents a fitted model.

### 2.2.3 Locating the membrane-bounded inclusion

Dialysis-bag positions were evaluated using the ABS-derived templates and the same procedure as for the ABS cylinder, so the reference and test objects differed. In the balanced 42-run cohort the position was correct in 33 of 42 runs at 160 kHz, and between 30 and 38 of 42 (71–90 %) at every frequency from 10 to 180 kHz, against a chance rate of 33.3 %; at 1 kHz accuracy fell to 17 of 42 (Figure 5a). Position recovery for the bags was lower than for the rigid insulator, consistent with their smaller perturbation magnitudes.

Position-classification errors were concentrated at the weakest conductivity contrasts. Accuracy tracks the contrast available to be seen (Figure 5b): 6 of 6 at 0.1× and 10×, 6 of 6 at 50×, 5 of 6 at 5× and 100×, but 3 of 6 at 0.5× and 2 of 6 at the matched condition. Six of the nine confusions at 160 kHz occur at those two weakest contrasts. Localization accuracy decreased with contrast and reached the chance level in the matched-conductivity condition, where no electrolyte conductivity contrast was available to localize. Thus, the matched bag produced a detectable overall perturbation, although its position was not consistently distinguished in these measurements (Section 2.2.2).

Because both inclusion types were measured at the same three positions, we could also ask which makes the better positional reference. Templates built from the ABS runs outperform templates built from the dialysis-bag runs, and do so even when the runs being classified are bags (33 of 42 against 27 of 42 at 160 kHz). The rigid insulator produces the strongest and most repeatable perturbation, so its average is the least noisy estimate of the pattern belonging to a position. A methodological argument points the same way: an ABS template is derived from a different object in different sessions and is therefore independent of the bag data it is used to interpret, whereas a bag-derived template shares its population with the runs being scored. A strong, rigid, repeatable inclusion is accordingly the better positional reference.

Two controls frame these numbers. Templates derived in 0.01 M background classified 9 of 12 bag runs acquired in 0.1 M background, providing initial evidence of positional-pattern transfer between backgrounds, with broader validation remaining for future work. Referencing every run to a single pooled empty-tank frame instead of a session-specific reference reduced accuracy; this pooled-reference comparison included the held-out session and is therefore presented descriptively rather than as independent validation. Template correlation is used throughout as the primary method because it is parameter-free and directly interpretable; a shrinkage linear discriminant analysis run under the identical protocol agreed within a few runs.

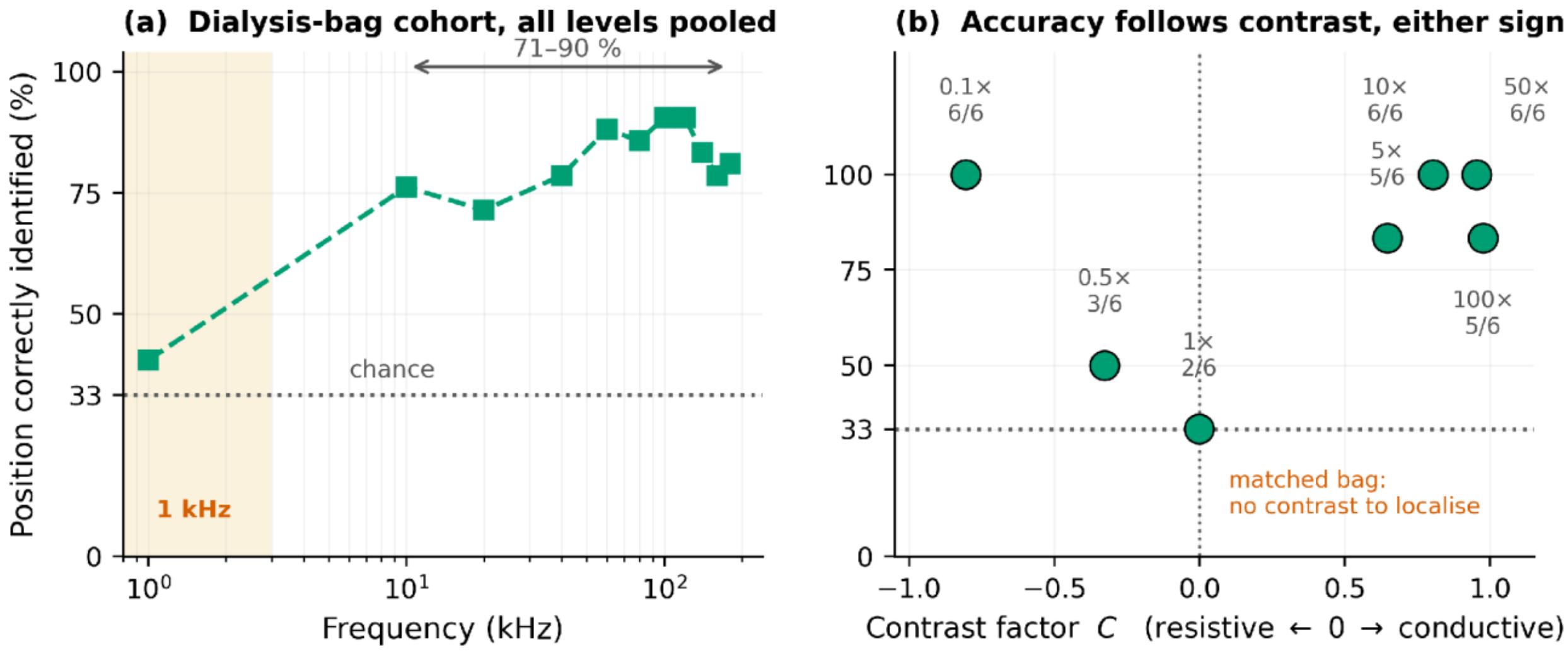

Figure 5. Locating the membrane-bounded inclusion. (a) Percentage of the 42 balanced dialysis-bag runs whose position was correctly identified, against excitation frequency, using ABS templates under leave-one-session-out validation. Accuracy is 71–90 % throughout 10–180 kHz and falls towards chance at 1 kHz (orange shading, excluded from analysis). (b) The same accuracy resolved by concentration level and plotted against the signed contrast factor C, with the count of correct runs beside each point. Accuracy is high wherever contrast is large, of either sign, and falls to chance at the matched condition where no conductivity contrast exists. The dashed line joins measured points and is a guide to the eye; the dotted lines mark C = 0 and the chance rate for three classes (33.3 %).

### 2.2.4 Position dependence and rotational symmetry

Perturbation magnitude did not differ systematically among the three angular positions (median RMS Δ 1.8 %, 2.2 % and 2.1 % for 1–16, 6–7 and 12–13 respectively, pooling both inclusion types at 160 kHz; all pairwise Welch comparisons non-significant), which is what the rotational symmetry of the electrode array predicts. The shape of the pattern, by contrast, is strongly position-dependent even where its magnitude is not: for the ABS inclusion the correlation of each run with the mean pattern of its own position (leave-one-run-out) was 0.60 ± 0.17, against 0.15 ± 0.15 for the two other positions (Welch $t = 11.9$, $p = 1.6 \times 10^{-16}$; n = 30 runs, 90 cross-correlations). The separation corresponds to Cohen's d = 2.8 and an area under the curve of 0.96. Magnitude therefore does not show where the inclusion is, and shape contains more information, which is why the classifier in Section 2.2.1 uses pattern correlation rather than amplitude.

A correlation of 0.60 should not be read as poor agreement. Although correlations are mathematically bounded by unity, the observed agreement is constrained by run-to-run variability: replicate frames acquired within a single session correlate at 0.92 ± 0.08, while independent runs of the same position on different days correlate at only 0.45 ± 0.22. Averaging several runs into a template recovers part of that loss, which is why the class-mean template outperforms any individual run used as a reference.

The tested positions were restricted to three angular locations. They were separated by 120° at a fixed radius of 4 cm (Section 5.14). The dataset contains no radial series; extending these observations to other depths, including central positions, will require additional measurements (Section 3.6).

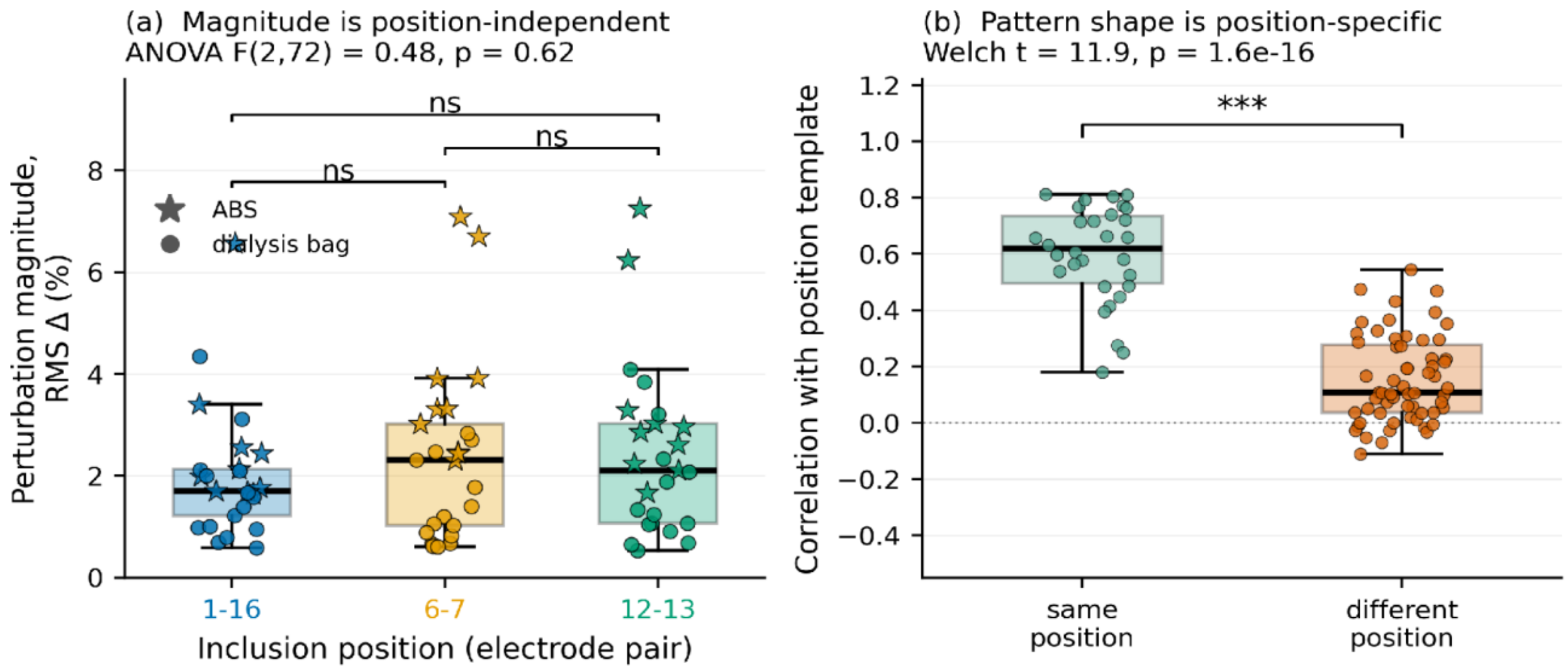

Figure 6. Position dependence at 160 kHz. (a) RMS Δ by angular position, shown as box plots with every run overlaid (stars, ABS; circles, dialysis bag); brackets give pairwise Welch comparisons, all non-significant. Magnitude is position-independent, consistent with the rotational symmetry of the array. (b) Mean leave-one-run-out Pearson correlation between each ABS run's perturbation pattern and the mean pattern of each position, shown as box plots with individual points and the Welch comparison bracketed: the pattern shape rotates with the inclusion.

## 3. Discussion

### 3.1 Arduino Giga as an EIT controller

The results demonstrate that the Arduino Giga R1 WiFi can serve as the complete signal-generation and data-acquisition controller for a functional multi-frequency EIT system operating across 1–180 kHz without dedicated analogue signal-generation hardware. The dual-core STM32H747XI architecture is well suited to this task: the M4 core's exclusive dedication to the waveform timing loop minimizes jitter, while the M7 core handles control and acquisition without competing for execution time.

Measured performance places the platform at 46–50 dB median boundary-voltage SNR across 10–180 kHz, against approximately 65–80 dB for purpose-built research instruments (Avery et al., 2017; Wi et al., 2014). Under these experimental conditions, inclusion position was recovered with 97 % accuracy at 160 kHz (29 of 30 runs in the balanced ABS cohort). Classification draws on the full 224-channel pattern; because channel noise is correlated (mean absolute off-diagonal correlation 0.33), these channels are not treated as independent measurements. Session-specific empty-tank referencing also helps manage the greater baseline variability observed across sessions (Section 3.6). The practical implication is that short-interval electronic noise was not the dominant limitation for these pattern-classification experiments.

Square-wave excitation is a deliberate simplification of the hardware and supported the present experiments in predominantly resistive electrolyte media, although harmonics were not separately quantified. For future studies requiring separate resistive and reactive components, sinusoidal excitation with phase-sensitive demodulation would be an appropriate extension, potentially using the onboard DAC and M7 resources.

### 3.2 Frequency dependence: electrode polarization and capacitive coupling

The frequency trends observed in the raw voltage statistics (Sections 2.2.1–2.2.2) may be influenced by electrode polarization. At frequencies below approximately 10 kHz, electrode polarizations impedance, arising from the double-layer capacitance at the silver wire–electrolyte interface, becomes comparable to or larger than the electrolyte resistance. Such interfacial impedance may affect the delivered current and measured boundary voltages. The Ag/AgCl surface layer that forms on silver wire in chloride media reduces this polarization impedance relative to bare metal, but does not eliminate it at the lowest frequencies studied.

Two independent measurements locate this transition. Boundary-voltage SNR falls from a 46–50 dB plateau across 10–180 kHz to 38 dB at 1 kHz. Independently, inclusion-position classification accuracy from raw voltage statistics is 87–97 % across 10–180 kHz for the ABS cohort, but falls to 30 % at 1 kHz, close to the 33 % chance level. Accordingly, 1 kHz was excluded from the main 10–180 kHz analyses but retained in the comparison plots.

Within the analysis band, the residual low-frequency excess perturbation (Section 2.2.2) decays monotonically to a high-frequency plateau and is largest where genuine contrast is weakest, suggesting that electrode-interface effects may contribute. Phase-resolved acquisition would help examine the separate contribution of membrane capacitance (Section 3.1).

Together, the observed signal quality and classification results support a 10–180 kHz working window, with high position-classification accuracy observed across that band. After session-baseline correction, accuracy is 87–97 % across the band for the ABS cohort. We therefore describe 10–180 kHz as the working window rather than a single optimum.

### 3.3 Information from boundary voltages

In the tested phantoms, three quantities were estimated directly from boundary voltages. First, inclusion position: matching a 224-channel perturbation pattern against previously recorded patterns identified the correct one of three positions in 29 of 30 runs (96.7 %; 95 % CI 83–100 %) at 160 kHz for the ABS cohort, and 33 of 42 for the balanced dialysis cohort, under leave-one-session-out validation in which all measurements from the test session were withheld, across independent sessions. Second, contrast polarity: the sign of the projection β onto a known insulating-inclusion pattern correctly identified whether the inclusion was more or less conductive than the background in all 36 dialysis-bag runs with genuine conductivity contrast in the balanced cohort (0.1×–0.5× all positive; 5×–100× all negative). Third, contrast magnitude, in the coarse sense described in Section 2.2.2, linear in the cylindrical contrast factor ($R^2 = 0.89$ for the aggregate fit shown in Figure 3) but saturating above roughly tenfold conductivity ratio, so that broad contrast classes rather than fine concentration steps are resolved. The present analysis focuses on relative conductivity contrast and three predefined angular positions rather than calibrated absolute conductivity, inclusion size, or shape. As expected, position discrimination was less distinct for bags with conductivity close to the background (Section 2.2.3).

This observation is of practical interest for applications where the full tomographic reconstruction is not feasible or necessary, and where a rapid, low-computation indicator of relative conductivity change is sufficient. A classification rule based on the correlation of the 224-channel relative-perturbation vector against a small set of stored reference patterns could in principle be computed on the Arduino Giga after frame acquisition, given that the operation is a single inner product per stored pattern, of order $10^3$ multiply-accumulate operations for a three-class decision, well within the M7 core's per-frame budget of several seconds.

Future work should investigate whether machine-learning classifiers trained on these features, or on richer feature vectors derived from the full frequency-dependent voltage profile, can improve discrimination performance across sessions, and at what level of experimental noise this advantage degrades. Further work can extend the measurement-domain approach beyond the relative contrast and predefined-position tasks evaluated here; position accuracy for the matched-conductivity bag was at the chance level in the present cohort. In this dataset, a multi-frequency feature vector concatenating all ten analysis frequencies yielded 84 % accuracy versus 87 % for the single-frequency rule at 160 kHz, suggesting value in testing richer features across a larger number of independent sessions. Separately, templates trained in 0.01 M background correctly classified inclusion position in 9 of 12 test measurements in 0.1 M background, suggesting partial transfer of positional information between the two backgrounds, with two of the three failures occurring at that series' matched condition.

### 3.4 Dialysis bag as a biologically motivated phantom

The choice to use regenerated cellulose dialysis bags reflects an interest in phantom media with membrane boundaries. Conventional EIT phantoms typically use rigid agar gel cylinders or silicone rods that present either a near-perfect conductor or near-perfect insulator boundary; the dialysis membrane provides an ion-permeable compartment boundary, offering a simplified analogue of selected tissue interfaces. A 12 kDa molecular-weight cut-off passes $K^+$ and $Cl^-$ essentially freely, so the bag is deliberately ion-permeable, a diffusion barrier and thin dielectric interface rather than a seal; this provides a membrane-bounded test inclusion (Section 5.12).

The pre-equilibration step described in Section 5.12, immersing the membrane in background electrolyte before filling, hydrates the membrane and conditions it for handling. It does not reduce the transmembrane concentration gradient once the bag is filled with a solution of different concentration, and therefore does not by itself limit diffusive exchange.

Ion exchange across the membrane was not directly quantified. The measured intercept of the contrast relationship is −0.031, and the matched-concentration bag produced a perturbation of ≈1.7 % RMS (median) against a 0.4–0.6 % noise floor; both are consistent with a residual bag-associated signal, potentially reflecting the membrane and displaced electrolyte. Ion exchange during a session may change the initial concentration contrast; direct measurements of time-dependent bag conductivity would help refine future quantitative calibration. The present fitted relationship is therefore interpreted as a controlled concentration-contrast trend rather than an independent conductivity calibration.

### 3.5 Electrolyte concentrations and physiological context

The KCl concentrations used here (0.001–1.0 M, measured conductivities 0.015–10.6 S/m at 22 °C, Supplementary Figure S3) span the conductivity range of the major biological body fluid compartments: cerebrospinal fluid (1.79 S/m at 37 °C; Baumann et al., 1997), blood plasma (approximately 1.4–1.6 S/m at body temperature, an essentially protein-loaded ~0.15 M saline; Hirsch et al., 1950), whole blood at low frequency (approximately 0.7 S/m, lower than plasma because erythrocyte membranes are insulating below the β-dispersion; Gabriel et al., 1996), and interstitial fluid, sweat and urine, whose ionic conductivities span roughly 0.1–3 S/m depending on composition and hydration (Grimnes and Martinsen, 2015). This selection means the phantom conductivities are numerically representative of the biological setting, at least in terms of bulk ionic conductivity magnitude, while extending above it at the top of the range.

These benchtop experiments establish performance in controlled electrolyte phantoms, where geometry and conductivity contrast can be varied systematically. Translating the approach to muscle and other soft tissues will require further evaluation of three-dimensional anatomy, skin–electrode contact, tissue heterogeneity, and motion. The overlap with physiological conductivity ranges provides motivation for these studies, while the present work makes no claim of in vivo validation or wearable performance.

The present findings motivate a staged progression from controlled electrolyte phantoms toward more complex phantoms and biological tissues. Although anatomically detailed phantoms approximate muscle, fat, bone and vasculature, they provide useful intermediate tests of geometry, electrode contact and tissue heterogeneity. An alternative is to establish, on simple phantoms of exactly known geometry, that a low-cost system recovers the quantities of interest — position, contrast polarity, coarse contrast magnitude — with characterized accuracy and a documented noise and drift budget, before progressing to tissue measurements with the necessary safety and validation controls. On this view the simple-phantom stage exists to characterize the instrument, not to simulate the subject.

Two developments would support this route. First, numerical simulation could complement phantom work: a finite-element forward model of the tank, validated against the measurements reported here, would allow inclusion geometry and conductivity distributions to be explored computationally, subject to experimental validation. Second, where anatomy is known approximately but not exactly, low-dimensional, compartment-based parameter estimation may offer a practical alternative to full conductivity mapping. Solving for a small number of conductivity values assigned to anatomically defined compartments could reduce the number of unknown parameters, consistent with what the present raw-voltage analysis indicates: the measurement supports a small number of robust parameters rather than a detailed image.

### 3.6 Limitations and future work

The present study demonstrates the feasibility of extracting spatial and conductivity-related information directly from boundary-voltage measurements using a low-cost, multi-frequency EIT platform. Several opportunities remain for further characterization and optimization. Direct measurement of injected current would enable more comprehensive evaluation of the current source and support future absolute-impedance measurements. Although the system exhibited good short-term measurement stability, session-to-session baseline variations highlight the importance of consistent acquisition conditions and reference measurements. Further improvements in hardware stability and signal processing may enhance measurement reproducibility.

The present experiments evaluated inclusion position and conductivity contrast under controlled conditions, using three angular positions at a fixed radius and a range of electrolyte concentrations. Expanding the experimental design to include additional inclusion sizes, radial positions, and independent measurement sessions would allow more comprehensive assessment of spatial sensitivity and generalizability. Future work will also explore phase-resolved measurements, advanced feature extraction, and more complex biological phantoms to evaluate the platform's potential for monitoring conductivity-related changes in soft tissues.

## 4. Conclusion

This study shows that an accessible, microcontroller-based EIT platform preserves structured information in its multichannel boundary measurements to determine location, contrast polarity, and coarse contrast magnitude without solving the inverse problem. We presented the design, implementation and characterization of a low-cost multi-frequency EIT system built around the Arduino Giga R1 WiFi microcontroller. The system uses the board's dual-core architecture to separate square-wave signal generation (M4 core, 1–180 kHz) from 16-bit ADC-based voltage acquisition with three-multiplexer electrode control (M7 core), supported by a discrete current source and passive RC injection and measurement filters.

The principal findings are: that the Arduino Giga, combined with three commodity multiplexer units and a simple analogue circuit, provides a functional multi-frequency EIT controller with no external signal-generation hardware, achieving a 46–50 dB median boundary-voltage SNR across 10–180 kHz; that raw boundary-voltage statistics, without any inverse solution, identify inclusion position in 29 of 30 runs of the insulating inclusion (96.7 % at 160 kHz, leave-one-session-out validation over multiple sessions) and 33 of 42 runs of the membrane-bounded inclusion, recover the polarity of the conductivity contrast in all 36 runs of the balanced cohort with genuine contrast, and yield a contrast magnitude linear in the cylindrical contrast factor with $R^2 = 0.89$ for the aggregate fit in Figure 3; that location recovery from raw statistics is essentially independent of excitation frequency between 10 and 180 kHz once session baselines are used (87–97 %), whereas at 1 kHz it falls to chance, highlighting the role of session-specific referencing in consistent performance across frequencies; and that the electrolyte concentrations studied span the conductivity range of major biological fluid compartments, providing a basis for future investigation in more complex geometries. The matched-conductivity membrane inclusion remains detectable above the noise floor, identifying the dialysis bag itself as an electrically non-trivial object and motivating phase-resolved follow-up.

Future work may assess targeted models that estimate a small number of conductivity-related parameters directly from boundary measurements.

**Acknowledgements**

We thank the members of the Center for Surgical Innovation and Engineering (CSIE) at Cedars-Sinai for their support. We additionally thank Prof. Tyler N. Tallman for technical discussions.

**Conflicts of interest**

The authors declare no conflict of interest.

## 5. Materials and Methods

### 5.1 Measurement Principles

Electrical impedance tomography (EIT) uses currents applied through boundary electrodes and voltage measurements on the same boundary to probe internal electrical properties. Here the measured boundary voltages are analysed directly, without calculating a conductivity image.

Sixteen electrodes were arranged around the tank. For each measurement cycle, current was applied through one adjacent electrode pair (e.g., electrodes 1 and 2) while voltage was sampled sequentially at the electrodes. The injection pair was then advanced (e.g., to electrodes 2 and 3) until all 16 adjacent pairs had been used.

### 5.2 System Overview and Physical Assembly

The EIT system was built around two primary sub-assemblies linked by a 16-wire electrode harness: a cylindrical measurement tank housing the electrode array, and a separate electronics enclosure. Enclosure construction and host connection are described in Supplementary Information S1.

The Arduino Giga R1 WiFi is a commercially manufactured development board built around the STM32H747XI, a dual-core system-on-chip integrating an ARM Cortex-M7 core (up to 480 MHz) and an ARM Cortex-M4 core (up to 240 MHz). The board provides 76 digital input/output pins, multiple analog input and output channels, hardware timers, and onboard 802.11 b/g/n WiFi and Bluetooth connectivity, all within the widely known Arduino programming environment.

### 5.3 Dual-Core Architecture

The Arduino Giga R1 WiFi is built on the STM32H747XI dual-core processor, and we exploited this architecture to cleanly separate signal generation from acquisition and control. The two cores communicated through a shared memory region managed by the Arduino RPC library and ran in parallel throughout operation.

Signal generation was assigned exclusively to the Cortex-M4 core and all other functions to the Cortex-M7 core; frequency selection and signal-generation control are described in Sections 5.4 and 5.9.

### 5.4 Signal Generation

The M4 core generated symmetric square-wave excitation at eleven discrete frequencies — 1, 10, 20, 40, 60, 80, 100, 120, 140, 160, and 180 kHz — spanning the range over which biological and electrochemical impedance spectra carry useful information. Frequency selection was issued by the M7 core over the inter-core GPIO bus. To ensure clean transitions, waveform generation was briefly paused before each frequency update and resumed only after the new half-period was set.

A symmetric square wave at fundamental $f_0$ carries odd harmonics at $3f_0$, $5f_0$, $7f_0$, … with amplitudes 1/3, 1/5, 1/7 of the fundamental. Because the measurement chain applies no frequency-selective demodulation (Section 5.8), the recorded value at a nominal frequency is the response to this periodic excitation as a whole rather than the impedance at a single frequency; the injection-path low-pass filter (Section 5.5) attenuates the higher harmonics progressively across the sweep. Frequency-resolved impedance in the strict sense would require sinusoidal excitation or synchronous (lock-in) demodulation at $f_0$, both feasible extensions of the platform (Section 3.1).

### 5.5 Analogue Circuit: Current Source, Filtering, and Measurement Front-End

The square-wave output from the M4 core drove a current pump, a topology chosen because it presents high output impedance and is intended to approximate constant-current injection over the operating range (Pease, 2008; Adler and Holder, 2021). The circuit was configured for nominal alternating-current injection, with a magnitude set by the resistor network and op-amp configuration.

Two passive RC filter stages were incorporated to manage noise without disturbing the signal across the operating band. A low-pass filter on the injection pathway sat between the current pump output and MUX 1, while a high-pass filter on the measurement pathway sat between MUX 3 and the ADC input. Resistor and capacitor values were selected for the target excitation band. Corner frequencies were deliberately placed outside the 1–180 kHz operating range to maintain a near-flat response throughout.

### 5.6 Three-Multiplexer Electrode Switching Architecture

Electrode switching was handled by three independent 16-channel analogue multiplexers, each assigned a dedicated role and controlled by the M7 core: MUX 1, positive current injection pathway; MUX 2, negative current return pathway; MUX 3, sequential voltage measurement pathway, connected to the ADC input.

MUX 1 and MUX 2 operated as a coordinated pair for current injection, while MUX 3 stepped through all 16 electrodes to acquire voltage. Before updating any injection configuration, waveform generation was paused to avoid transient artefacts; measurements began only after a short settling delay (10 μs before data acquisition; 4 μs per MUX step).

### 5.7 Injection Pattern and Data Organization

Current was injected between adjacent electrode pairs following a standard neighboring pattern. To mitigate residual charge accumulation at the electrode double-layer capacitance, injection pairs were addressed in an interleaved sequence so that no two consecutive injections shared an electrode. The interleaving introduced a non-sequential ordering in the raw data, which was corrected by a post-processing function, sort_arduino_data(), before any analysis was performed.

At each excitation frequency, the system acquired a 256-element voltage vector — 16 injection pairs, each measured across all 16 electrode positions — giving a complete dataset of 11 × 256 voltages per acquisition, transmitted as a single matrix over USB serial. Of the 256 channels, 224 are measurements on non-driven electrodes (16 pairs × 14); the 32 driven-electrode channels are dominated by contact impedance and are excluded from the classifier features (Section 5.16). The adjacent-drive pattern is well suited to the present near-boundary phantom positions (r = 4 cm in a 14.9 cm bore); sensitivity at greater depths can be examined with alternative injection patterns in future experiments.

### 5.8 Voltage Averaging and Threshold Gating

To improve measurement stability, each electrode voltage was averaged over a two-level loop: an outer loop of n_avg = 100 iterations, each accumulating n_avg_each = 100 ADC samples, yielding 10,000 raw readings per electrode per injection pair per frequency. A configurable threshold gate was applied to reject near-zero samples associated with incomplete signal stabilization; this gate was deliberately disabled for the current-injecting electrodes, where near-zero differential voltages are expected. Accepted samples were averaged and scaled using the ADC reference and the 16-bit full-scale count (65,535) to produce voltages in volts. The reported quantity per channel is therefore a gated mean of the single-ended waveform, an amplitude-like statistic of the periodic response, rather than a synchronously demodulated impedance magnitude (Section 5.4).

The sample budget is consistent with the measured acquisition time: 16 injection pairs × 16 electrodes × 100 × 100 = $2.56 \times 10^6$ conversions per frequency, and $2.82 \times 10^7$ per 11-frequency sweep; over the ≈60 s sweep this is a sustained rate near $4.7 \times 10^5$ conversions per second, comfortably within the STM32H747's ADC capability and leaving margin for multiplexer settling, inter-core handshaking, and serial transfer.

### 5.9 Grounding Sequence and Acquisition Protocol

Before each measurement sweep, we performed a grounding procedure in which MUX 2 cycled sequentially through all electrodes while the current source remained inactive (100 cycles × 10 ms per electrode). This step was intended to discharge residual charge at the electrode-electrolyte interfaces and improve the consistency of measurements across sweeps.

### 5.10 Tank and Electrode Array

The measurement tank was printed in black PETG by fused deposition modelling at a 0.2 mm layer height. It had an outer diameter of 15.5 cm, a wall thickness of 3 mm, and a height of 50 mm; electrolyte fill height was maintained at 25 mm in all experiments. Sixteen half-round silver wire electrodes (length 6 cm; nominal full-round diameter 2.05 mm; 92.5% Ag) were equally spaced around the inner wall at 22.5° intervals. Each electrode was bent into a hook geometry and passed through a dedicated wall opening and secured into a notch to maintain consistent electrolyte contact.

Silver was chosen for its high electrical conductivity and because immersion in chloride-containing electrolytes rapidly forms a stable Ag/AgCl surface layer, reducing electrode polarization effects (Grimnes and Martinsen, 2015). The wire is sterling silver (92.5 % Ag, balance predominantly Cu) rather than fine silver, a choice made on cost and availability grounds consistent with the aims of the platform; the AgCl layer forms on the silver fraction, and the session-resolved bath conductivity measurements (Section 5.11) provide the evidence that copper-related contamination did not materially affect the electrolyte.

### 5.10.1 Electrode preparation and preconditioning

Electrodes were rinsed with deionized water and then preconditioned in two stages: they were first immersed in the working KCl electrolyte to allow the electrode–electrolyte interface to equilibrate, and were subsequently exercised through complete acquisition cycles, spanning the full eleven-frequency sweep and all injection patterns, before any data used in the analyses reported here were recorded. No deliberate chloridization step was applied; the Ag/AgCl surface layer formed in situ during immersion and during these preconditioning cycles. Preconditioning is reported because, in its absence, the electrode half-cell potential and polarization impedance evolve during the early part of a session and differ between sessions, which would confound both the frequency sweep and the session-grouped classification described in Section 5.16. The adequacy of the procedure was assessed by the repeated empty-tank frame statistics of Section 2.1 (0.4–0.6 % consecutive-frame RMS $\Delta V/V$ in stable series).

## 5.11 Reagents and Solution Preparation

All potassium chloride solutions were prepared from ACS reagent-grade KCl (99.0–100.5 % ACS analytical reagent grade, VWR Chemicals BDH®) dissolved in Type 1 ultrapure water (resistivity ≥18 MΩ·cm) and made up volumetrically to nominal concentrations of 0.001, 0.005, 0.01, 0.05, 0.1, 0.5, and 1.0 M for inclusion solutions, and 0.01 M and 0.1 M for background electrolytes. Electrical conductivity was measured at room temperature (22 ± 1°C) using a calibrated dual-channel benchtop meter (SevenDirect SD23, Mettler Toledo, Columbus, OH, USA) equipped with an InLab® 731-ISM conductivity/temperature sensor. Measured conductivity values, rather than nominal electrolyte concentrations, were used as ground-truth values throughout the study. Session-resolved measurements are summarized in Supplementary Figure S3 (36 inclusion-solution and 10 background measurements across nine sessions); the 0.01 M background measured 1323–1462 µS/cm (mean 1391 µS/cm, 0.139 S/m) across sessions.

## 5.12 Dialysis Membrane Preparation

Regenerated cellulose dialysis tubing (MWCO 12 kDa; flat width 44 mm) was cut to 120 mm lengths and soaked in 0.01 M KCl for at least 30 minutes to ensure full hydration, then thoroughly rinsed. Each section was subsequently equilibrated in the relevant background electrolyte for 15 minutes immediately before filling; this step conditioned the membrane in the background solution before introducing the inclusion electrolyte, without eliminating the concentration gradient that develops after filling. After filling with the inclusion solution, one end of the dialysis bag was secured with a nut positioned at the bottom of the tank, while the top end was sealed using a dialysis clip. The exterior of the bag was then lightly blotted dry before placing it into the tank.

Regenerated cellulose at a 12 kDa cut-off was chosen deliberately: the membrane passes $K^+$ and $Cl^-$ essentially freely, so the bag is an ion-permeable diffusion barrier and thin dielectric interface rather than a seal. That is the property being studied — a semi-permeable compartment boundary of the kind present in tissue (Section 3.4). Each filled bag spanned the full 25 mm electrolyte depth, preserving the quasi-two-dimensional geometry of the domain.

### 5.13 ABS Inclusion

An electrically insulating inclusion was 3D-printed in ABS at 100 % infill and 0.2 mm layer height. The structure was a hollow cylinder (outer diameter 30 mm, wall thickness 2 mm, height 55 mm). At 55 mm the inclusion extends well above the 25 mm fill height, so it spans the full depth of the electrolyte, and the domain is genuinely quasi-2D over the measurement plane.

### 5.14 Experimental Protocol

Three inclusion positions were defined symmetrically within the tank, each at a radial distance of r = 4 cm from the center and separated by 120°, adjacent respectively to electrode pairs 1–16, 6–7 and 12–13, by which they are labelled. For each experiment, the tank was filled with the selected KCl background electrolyte, and bulk conductivity was measured before positioning the inclusion. EIT measurements were then acquired across all eleven frequencies. Background conductivity was re-measured after each acquisition to monitor any drift over the course of the experiment. All experiments were conducted at room temperature (22 ± 1°C), and raw voltage data were stored as labelled CSV files for subsequent analysis.

Measurements were acquired over a number of experimental sessions; a session being defined as a single tank fill with a continuous run of acquisitions on one day; sessions are the grouping variable for the cross-validation of Section 5.16. The analyzed dataset comprises ten orientation-consistent sessions containing 75 inclusion runs in 0.01 M background — 30 ABS runs (62 frames including replicates) and 45 dialysis-bag runs across seven concentrations and three positions — plus one further session of 12 dialysis-bag runs in 0.1 M background used for the cross-electrolyte transfer test, 25 empty-tank time-lapse frames for noise and drift characterization, and the session empty-tank baselines.

### 5.15 Control Measurements

Measurement noise was quantified as a repeat-frame signal-to-noise ratio: 25 repeated empty-tank frames (three time-lapse series on two days) were acquired without altering the phantom, and for each channel the SNR was computed as the ratio of the mean channel value to its standard deviation across repeats, expressed in decibels; the standard deviation was estimated from first differences of consecutive frames (divided by $\sqrt{2}$) to remove slow drift, computed per channel and then summarized by the median over the 224 usable channels. Repeat-frame SNR was 46–50 dB across 10–180 kHz, with a pooled median of 48 dB. The 1 kHz point is excluded from this range because boundary-voltage SNR falls to 38 dB there and ABS inclusion-position classification falls to 30 %, near the 33 % chance level, consistent with electrode polarization impedance becoming comparable to the electrolyte resistance below approximately 10 kHz (Section 3.2). The primary analyses use 10–180 kHz, with 1 kHz retained for comparison in the frequency-dependent results.

$$SNR(k,e) = 20\, log_{10}\left(\frac{|\bar{V}(k,e)|}{s(k,e)}\right) \quad \text{Eq. 1}$$

Here V̄(k, e) denotes the mean over N repeated frames, and s(k, e) is estimated from differences between consecutive frames, reducing sensitivity to slow baseline drift:

$$s(k,e) = \sqrt{\frac{1}{2(N-1)} \sum_{i=1}^{N-1} (V_{i+1}(k,e) - V_i(k,e))^2} \quad \text{Eq. 2}$$

### 5.16 Boundary-Voltage Analysis and Classification

Baseline mapping. Each inclusion run is referenced to the nearest empty-tank frame acquired earlier in the same session, not to the mean of that session's empty frames. Where multiple empty frames were available, the nearest preceding frame was used as reference.

Primary classifier: nearest-class-mean template correlation. For each position l, the template T_l is the mean of Δ over training runs at that position; a test run is assigned to argmax_l |rho(Delta, T_l)|, where rho is the Pearson correlation over the 224 retained channels. Absolute value is used because inclusions of opposite contrast polarity produce mirrored patterns at the same position. Templates are built from ABS runs only, which give the strongest and most repeatable perturbation and are independent of the dialysis-bag data they are used to interpret. Validation is leave-one-session-out: every run acquired in the test run's session is withheld from template construction, so no measurement contributes to the template it is scored against.

Sensitivity analyses. Two alternatives were run under the identical split and identical training set: true 1-nearest-neighbour, in which a test run takes the label of the single most correlated training run rather than of the closest class mean; and shrinkage linear discriminant analysis with Ledoit-Wolf shrinkage. Sign alignment required by the discriminant model was performed without reference to the true label. These analyses were treated as sensitivity checks.

Analysis cohorts. The primary ABS cohort is balanced at ten runs per position (30 runs from multiple sessions). The primary dialysis cohort is balanced at six runs per concentration ratio (42 runs); the 0.1× level had nine runs available; one additional triplet was excluded from the balanced cohort because its measured conductivity changed during acquisition. The primary results use the 30-run ABS and 42-run balanced dialysis cohorts.

Each acquisition yields an 11 × 256 voltage matrix. Channels in which the measurement electrode coincides with either driving electrode were excluded (32 of 256), giving 224 usable channels per frequency. Two statistics are used throughout. The relative boundary-voltage perturbation of injection pair k and measurement electrode e is

$$\Delta(k,e) = \frac{V_{obj}(k,e) - V_{ref}(k,e)}{V_{ref}(k,e)} \quad \text{Eq. 3}$$

where Vref is the designated empty-tank baseline of the same session (the nearest preceding empty-tank frame from that session), which constitutes the session-specific drift correction; no further transformation (log scaling, standardization, or frequency-wise normalization) was applied. Perturbation magnitude is the RMS of Δ over the 224 usable channels,

$$RMS\,\Delta \;=\; \sqrt{\frac{1}{224}\sum_{(k,e)} \Delta(k,e)^2} \qquad \text{Eq. 4}$$

and the signed contrast projection is the inner product

$$\beta \;=\; \langle \Delta, \hat{t}_p \rangle \qquad \text{Eq. 5}$$

where $\hat{t}_p$ is the unit-normalized mean perturbation pattern of the insulating (ABS) inclusion at the same position p, so that $\beta > 0$ indicates a less-conductive inclusion and $\beta < 0$ a more-conductive one.

Inclusion position (three classes) was classified using the absolute Pearson correlation of the 224-channel perturbation vector with the three ABS-derived position templates. Templates were constructed from training sessions only, and performance was evaluated using leave-one-session-out validation. The expected chance accuracy was 33.3 %. An exploratory comparator used a pooled empty-tank baseline in place of the session-specific baseline. Concentration ranking was evaluated using Spearman correlation between $-\beta$ and the nominal concentration ratio. Cross-electrolyte transfer tests applied templates from 0.01 M KCl to runs acquired in 0.1 M KCl. For three exploratory ABS runs outside the trained angular positions, position was estimated using the amplitude-weighted circular mean of the injection-pair indices. Analyses were performed in Python using NumPy and SciPy.

## References

Adler, A., Holder, D. (Eds.) (2021). Electrical Impedance Tomography: Methods, History and Applications, 2nd ed. CRC Press, Boca Raton, FL. ISBN 978-1-032-16117-4.

Adler, A., Lionheart, W.R.B. (2006). Uses and abuses of EIDORS: an extensible software base for EIT. Physiological Measurement 27(5), S25–S42. doi:10.1088/0967-3334/27/5/S03

Aller, M., Mera, D., Cotos, J.M., Villaroya, S. (2023). Study and comparison of different machine learning-based approaches to solve the inverse problem in electrical impedance tomographies. Neural Computing and Applications 35(7), 5465–5477. doi:10.1007/s00521-022-07988-7

Avery, J., Dowrick, T., Faulkner, M., Goren, N., Holder, D. (2017). A versatile and reproducible multi-frequency electrical impedance tomography system. Sensors 17(2), 280. doi:10.3390/s17020280

Baumann, S.B., Wozny, D.R., Kelly, S.K., Meno, F.M. (1997). The electrical conductivity of human cerebrospinal fluid at body temperature. IEEE Transactions on Biomedical Engineering 44(3), 220–223. doi:10.1109/10.554770

Biasi, N., Gargano, A., Arcarisi, L., Carbonaro, N., Tognetti, A. (2022). Physics-based simulation and machine learning for the practical implementation of EIT-based tactile sensors. IEEE Sensors Journal 22(5), 4186–4196. doi:10.1109/JSEN.2022.3144038

Borcea, L. (2002). Electrical impedance tomography. Inverse Problems 18(6), R99–R136. doi:10.1088/0266-5611/18/6/201

Borsic, A., Attardo, E.A., Halter, R.J. (2012). Multi-GPU Jacobian accelerated computing for soft-field tomography. Physiological Measurement 33(10), 1703–1715. doi:10.1088/0967-3334/33/10/1703

Brazey, B., Haddab, Y., Zemiti, N. (2022). Robust imaging using electrical impedance tomography: review of current tools. Proceedings of the Royal Society A 478(2258), 20210713. doi:10.1098/rspa.2021.0713

Dunne, E., Santorelli, A., McGinley, B., Leader, G., O'Halloran, M., Porter, E. (2018). Image-based classification of bladder state using electrical impedance tomography. Physiological Measurement 39(12), 124001. doi:10.1088/1361-6579/aae6ed

Frerichs, I., Amato, M.B.P., van Kaam, A.H., Tingay, D.G., Zhao, Z., Grychtol, B., et al. (2017). Chest electrical impedance tomography examination, data analysis, terminology, clinical use and recommendations: consensus statement of the TRanslational EIT developmeNt stuDy group. Thorax 72(1), 83–93. doi:10.1136/thoraxjnl-2016-208357

Gabriel, S., Lau, R.W., Gabriel, C. (1996). The dielectric properties of biological tissues: II. Measurements in the frequency range 10 Hz to 20 GHz. Physics in Medicine and Biology 41(11), 2251–2269. doi:10.1088/0031-9155/41/11/002

Ghaednia, H., Owens, C.E., Roberts, R., Tallman, T.N., Hart, A.J., Varadarajan, K.M. (2020). Interfacial load monitoring and failure detection in total joint replacements via piezoresistive bone cement and electrical impedance tomography. Smart Materials and Structures 29(8), 085039. doi:10.1088/1361-665X/ab9e18

Grimnes, S., Martinsen, Ø.G. (2015). Bioimpedance and Bioelectricity Basics, 3rd ed. Academic Press, Oxford.

Grossi, M., Riccò, B. (2017). Electrical impedance spectroscopy (EIS) for biological analysis and food characterization: a review. Journal of Sensors and Sensor Systems 6(2), 303–325. doi:10.5194/jsss-6-303-2017

Hirsch, F.G., Texter, E.C., Wood, L.A., Ballard, W.C., Horan, F.E., Wright, I.S., et al. (1950). The electrical conductivity of blood: I. Relationship to erythrocyte concentration. Blood 5(11), 1017–1035. doi:10.1182/blood.V5.11.1017.1017

Ikuno, T., Kaneko, R. (2026). Task-oriented inference framework for lightweight and energy-efficient object localization in electrical impedance tomography. Sensors 26(8), 2570. doi:10.3390/s26082570

Keiderling, L., Rosendorf, J., Owens, C.E., Varadarajan, K.M., Hart, A.J., Schwab, J., Tallman, T.N., Ghaednia, H. (2023). Comparing machine learning algorithms for non-invasive detection and classification of failure in piezoresistive bone cement via electrical impedance tomography. Review of Scientific Instruments 94(12), 124103. doi:10.1063/5.0131671

Kim, K., et al. (2024). Extremely durable electrical impedance tomography–based soft and ultrathin wearable e-skin for three-dimensional tactile interfaces. Science Advances 10(38), eadr1099. doi:10.1126/sciadv.adr1099

Li, S., Kawashima, D., Dai, Z., Aoki, N., Takei, M. (2025). Electrical impedance tomography (EIT)-based intracellular conductivity imaging for non-invasive cell detection. Lab on a Chip 25(22), 5914–5925. doi:10.1039/D5LC00466G

Lionheart, W.R.B. (2004). EIT reconstruction algorithms: pitfalls, challenges and recent developments. Physiological Measurement 25(1), 125–142. doi:10.1088/0967-3334/25/1/021

Loke, M.H., Chambers, J.E., Rucker, D.F., Kuras, O., Wilkinson, P.B. (2013). Recent developments in the direct-current geoelectrical imaging method. Journal of Applied Geophysics 95, 135–156. doi:10.1016/j.jappgeo.2013.02.017

McDermott, B., O'Halloran, M., Porter, E., Santorelli, A. (2018). Brain haemorrhage detection using a SVM classifier with electrical impedance tomography measurement frames. PLOS ONE 13(7), e0200469. doi:10.1371/journal.pone.0200469

McDermott, B., Elahi, A., Santorelli, A., O'Halloran, M., Avery, J., Porter, E. (2020). Multi-frequency symmetry difference electrical impedance tomography with machine learning for human stroke diagnosis. Physiological Measurement 41(7), 075010. doi:10.1088/1361-6579/ab9e54

Oh, T.I., Wi, H., Kim, D.Y., Yoo, P.J., Woo, E.J. (2011). A fully parallel multi-frequency EIT system with flexible electrode configuration: KHU Mark2. Physiological Measurement 32(7), 835–849. doi:10.1088/0967-3334/32/7/S08

Olson, G., Hansmann-Canas, I., Karimi, Z., Yazdkhasti, A., Shabestanipour, G., Ghaednia, H., Schwab, J.H. (2025). The impact of AI on the development of multimodal wearable devices in musculoskeletal medicine. HSS Journal 21(3), 314–324. doi:10.1177/15563316251344945

Pease, R.A. (2008). AN-1515: A Comprehensive Study of the Howland Current Pump. Texas Instruments Application Report SNOA474. Texas Instruments, Dallas, TX.

Pennati, F., Angelucci, A., Morelli, L., Bardini, S., Barzanti, E., Cavallini, F., Conelli, A., Di Federico, G., Paganelli, C., Aliverti, A. (2023). Electrical impedance tomography: from the traditional design to the novel frontier of wearables. Sensors 23(3), 1182. doi:10.3390/s23031182

Rintoul, J. (2018). OpenEIT [computer software]. https://github.com/OpenEIT/OpenEIT (accessed August 2026).

Rymarczyk, T., Kłosowski, G., Kozłowski, E., Tchórzewski, P. (2019). Comparison of selected machine learning algorithms for industrial electrical tomography. Sensors 19(7), 1521. doi:10.3390/s19071521

Sciospec Scientific Instruments GmbH (n.d.). ISX-3 impedance analyser. Bennewitz, Germany. https://www.sciospec.com (accessed August 2026).

Shi, Y., Yang, Z., Xie, F., Ren, S., Xu, S. (2021). The research progress of electrical impedance tomography for lung monitoring. Frontiers in Bioengineering and Biotechnology 9, 726652. doi:10.3389/fbioe.2021.726652

Vahabi, N., Yerworth, R., Miedema, M., van Kaam, A., Bayford, R., Demosthenous, A. (2021). Deep analysis of EIT dataset to classify apnea and non-apnea cases in neonatal patients. IEEE Access 9, 25131–25139. doi:10.1109/ACCESS.2021.3056558

Wi, H., Sohal, H., McEwan, A.L., Woo, E.J., Oh, T.I. (2014). Multi-frequency electrical impedance tomography system with automatic self-calibration for long-term monitoring. IEEE Transactions on Biomedical Circuits and Systems 8(1), 119–128. doi:10.1109/TBCAS.2013.2256785

Yao, J., Takei, M. (2017). Application of process tomography to multiphase flow measurement in industrial and biomedical fields: a review. IEEE Sensors Journal 17(24), 8196–8205. doi:10.1109/JSEN.2017.2682929

Zheng, J., Cao, H., Chen, D., Ansari, R., Chen, K.-C., et al. (2022). A review of EMG-, FMG-, and EIT-based biosensors and relevant human–machine interactivities and biomedical applications. Biosensors 12(7), 516. doi:10.3390/bios12070516

Zhu, J., Snowden, J.C., Verdejo, J., Chen, E., Zhang, P., Ghaednia, H., Schwab, J.H., Mueller, S. (2021). EIT-kit: an electrical impedance tomography toolkit for health and motion sensing. In: Proceedings of the 34th Annual ACM Symposium on User Interface Software and Technology (UIST '21). ACM, New York. doi:10.1145/3472749.3474758

**Appendix A. Nomenclature**

Symbols are listed in the order in which they are introduced. Units are given where a symbol carries one; dimensionless quantities are identified as such. Note that σ denotes conductivity throughout this work and is never used for a standard deviation, which is written s.

| Symbol | Meaning, units, and where defined |
|---|---|
| $k$ | Index of the current-injection electrode pair, 1–16; pair k drives electrodes k and k+1. |
| $e$ | Index of the electrode at which potential is measured, 1–16. |
| $(k, e)$ | A measurement channel. A frame contains 256 channels; the 32 with e = k or e = k+1 are discarded, leaving 224. |
| $V$ | Boundary voltage frame, in volts; $V \in \mathbb{R}^{11\times256}$ for a complete eleven-frequency sweep. |
| $V_{obj}$ | Boundary voltage recorded with the inclusion present (V). |
| $V_{ref}$ | Empty-tank boundary voltage from the nearest preceding reference frame of the same session (V). Eq. 3. |
| $\Delta(k, e)$ | Relative perturbation, dimensionless, reported as a percentage: $\Delta = (V_{obj} - V_{ref})/V_{ref}$. Eq. 3. |
| $RMS\ \Delta$ | Root-mean-square of Δ over the 224 retained channels; the scalar measure of perturbation magnitude (%). Eq. 4. |
| $p$ | Angular position label of the inclusion: 1–16, 6–7 or 12–13, at radius 4 cm and 120° apart. |
| $\hat{t}_p$ | Unit-normalized mean perturbation pattern of the training runs at position p, i.e. the template. Built from ABS runs only. Eq. 5. |
| $\beta$ | Signed contrast projection, dimensionless: the inner product of Δ with $\hat{t}p$. $\beta > 0$ denotes an inclusion less conductive than the background, $\beta < 0$ more conductive. Eq. 5. |
| $\rho$ | Pearson product-moment correlation coefficient. Position is assigned by the largest $\lvert\rho\rvert$ between Δ and the three templates. |
| $\sigma$ | Electrical conductivity (S $m^{-1}$, or µS $cm^{-1}$ where stated). Never used for a standard deviation in this work. |
| $\sigma_{inc}$, $\sigma_{bkg}$ | Conductivity of the inclusion contents and of the background electrolyte respectively. |
| $C$ | Cylindrical contrast factor, dimensionless: $C = (\sigma_{inc} - \sigma_{bkg})/(\sigma_{inc} + \sigma_{bkg})$. C → +1 for a perfect conductor, −1 for a perfect insulator, and 0 at the matched condition. |
| $SNR(k, e)$ | Per-channel signal-to-noise ratio, in decibels. Eq. 1. |
| $\bar{V}(k, e)$ | Mean boundary voltage on channel (k, e) over the N repeated frames (V). Eq. 1. |
| $s(k, e)$ | Standard deviation of channel (k, e), computed from the differences between consecutive frames (V). Eq. 2. |
| $N$ | Number of repeated empty-tank frames in the noise analysis (N = 25). Eqs. 1–2. |
| $V_i(k, e)$ | Reading on channel (k, e) in the i-th frame of a repeated series (V). Eq. 2. |
| $n$ | Number of runs in a cohort: n = 30 for the ABS cohort, n = 42 for the balanced dialysis cohort. |
| $F(df_1, df_2)$ | Fisher F statistic of a one-way analysis of variance, with numerator and denominator degrees of freedom. |
| $R^2$ | Coefficient of determination of a least-squares fit. |

## Supplementary Information

### S1 System enclosure and host connection

The electronics enclosure houses the Arduino Giga, the current-source and signal-conditioning circuit, and the three multiplexers, linked to the tank by a 16-wire electrode harness; the host PC connects over USB serial and is used only for configuration and offline processing.

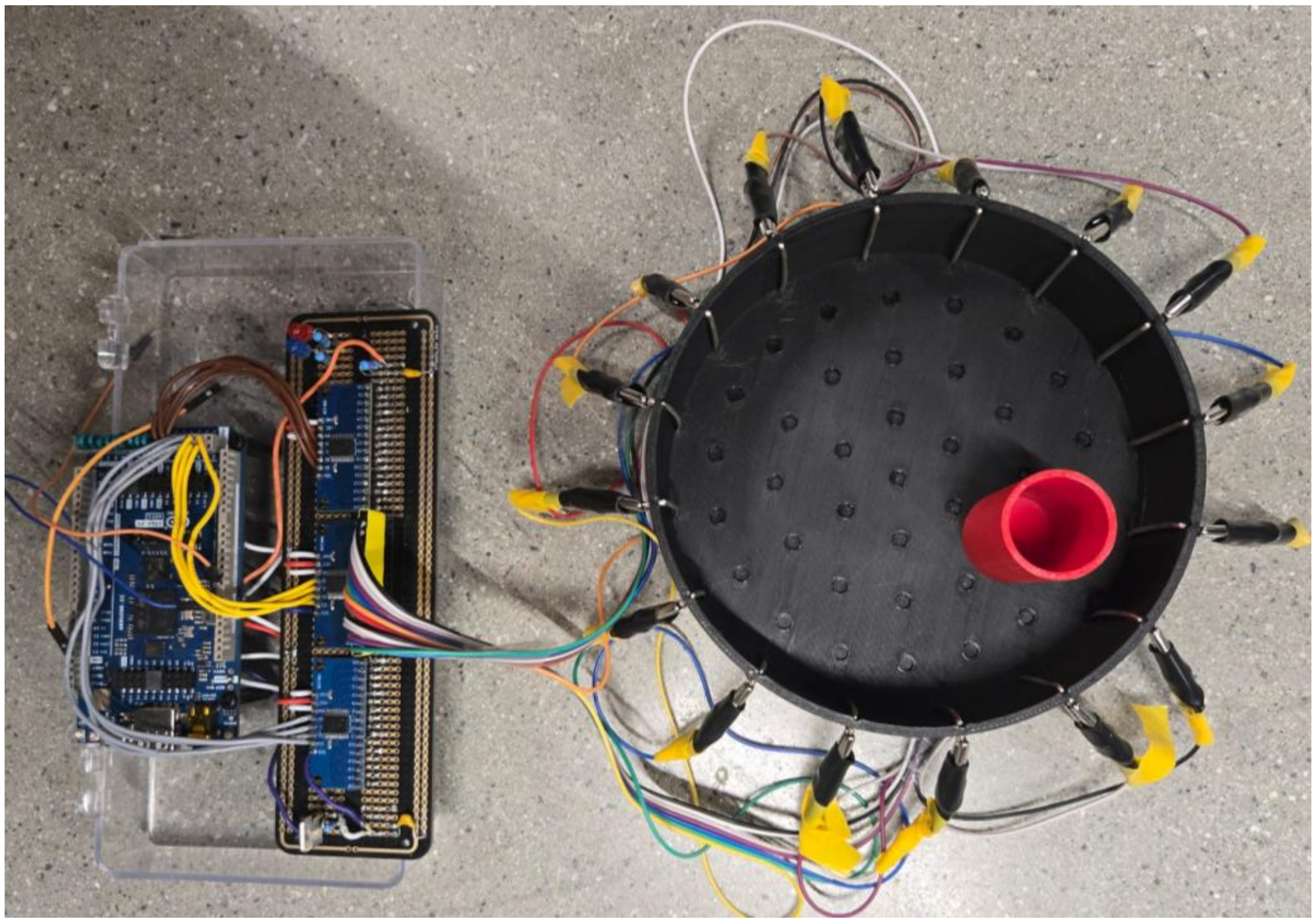

Supplementary photograph of the benchtop EIT system, showing the 16-electrode cylindrical tank, a representative central inclusion, and the separate electronics enclosure connected by the electrode harness.

### S2 Supplementary figures and supporting measurements

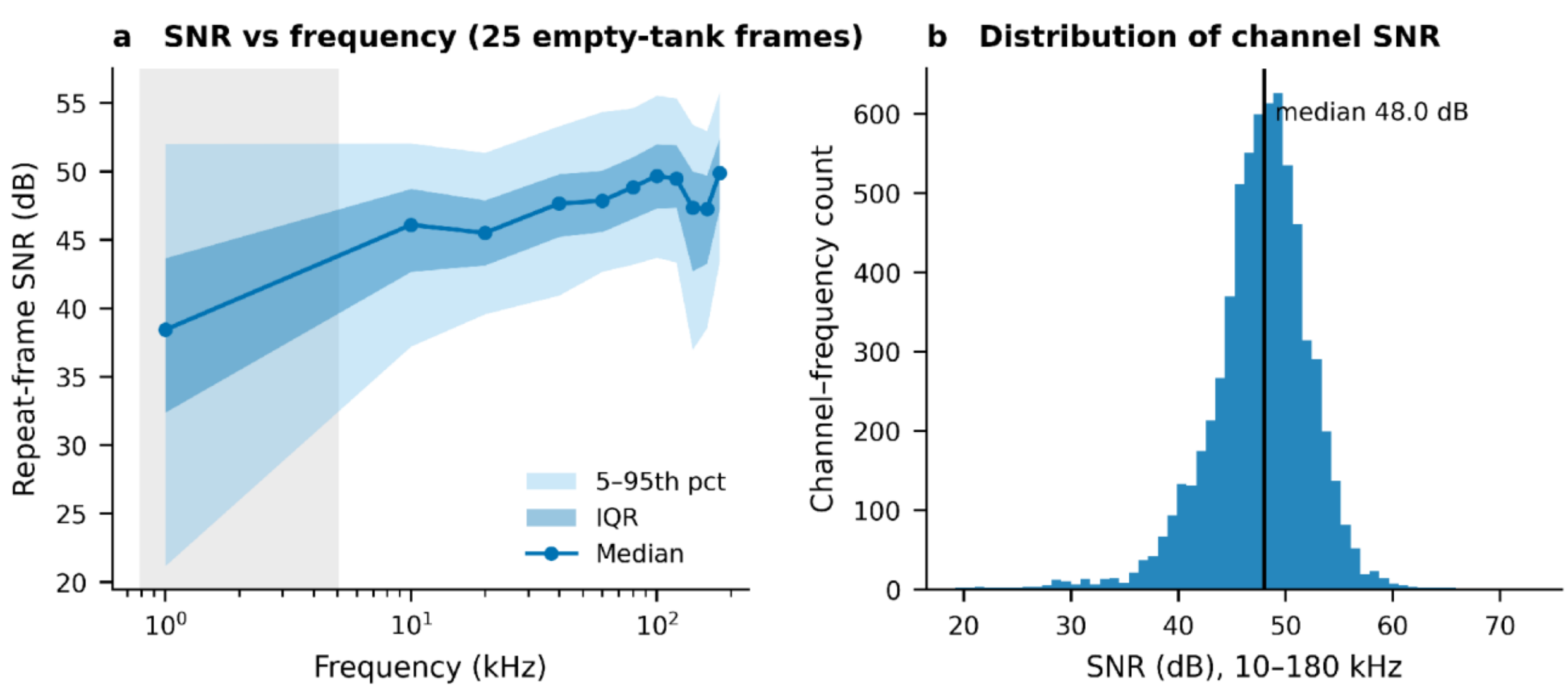

**Figure S1.** Per-channel repeat-frame signal-to-noise ratio, from 25 repeated empty-tank frames (0.01 M KCl; three time-lapse series on two days), with the standard deviation estimated from first differences of consecutive frames to remove slow drift. (a) SNR versus excitation frequency: median, interquartile range and 5–95th percentile across the 224 usable channels; the shaded band marks the excluded 1 kHz measurement. (b) Distribution of all 6720 channel–frequency SNR estimates from 10 to 180 kHz; the line marks the 48 dB median.

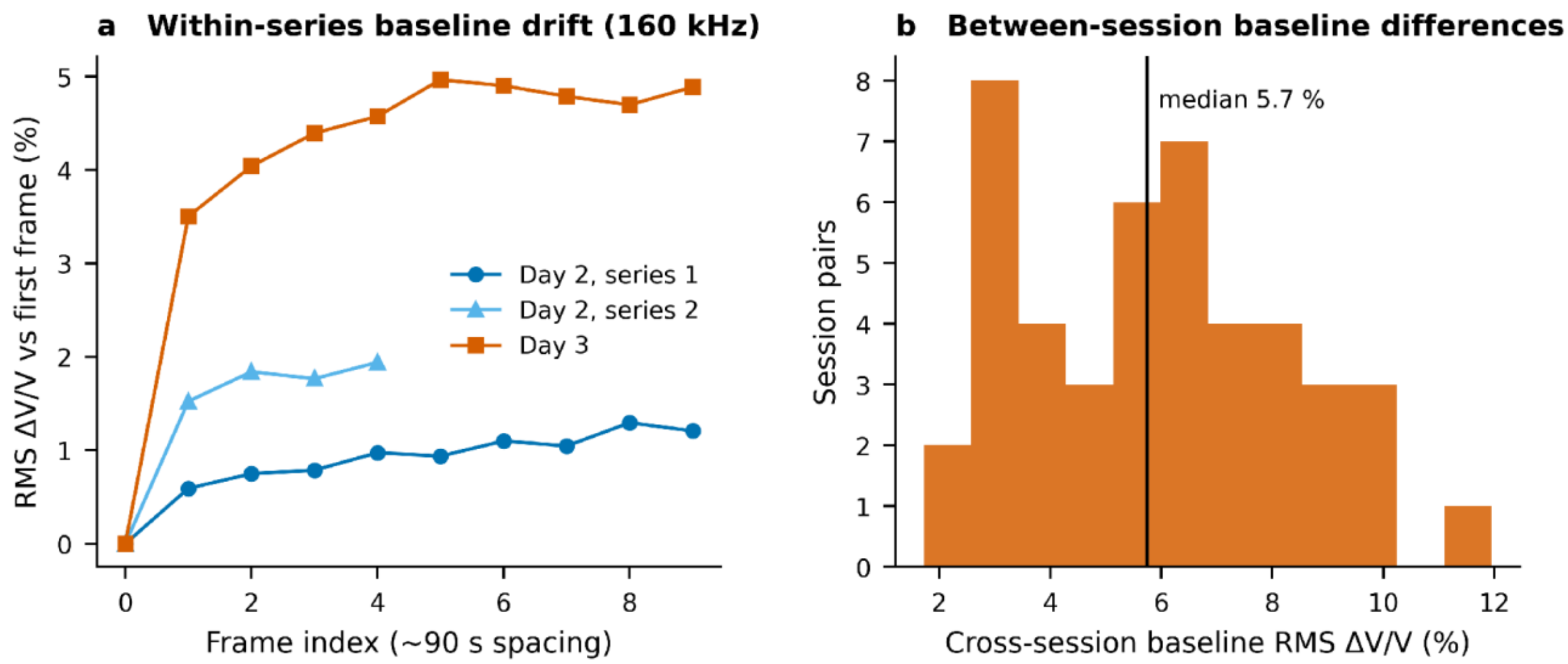

Figure S2. Baseline stability at 160 kHz. (a) RMS ΔV/V of each empty-tank time-lapse frame relative to the first frame of its series (≈90 s frame spacing). (b) Distribution of pairwise between-session empty-tank baseline differences (45 session pairs; median 5.7 %).

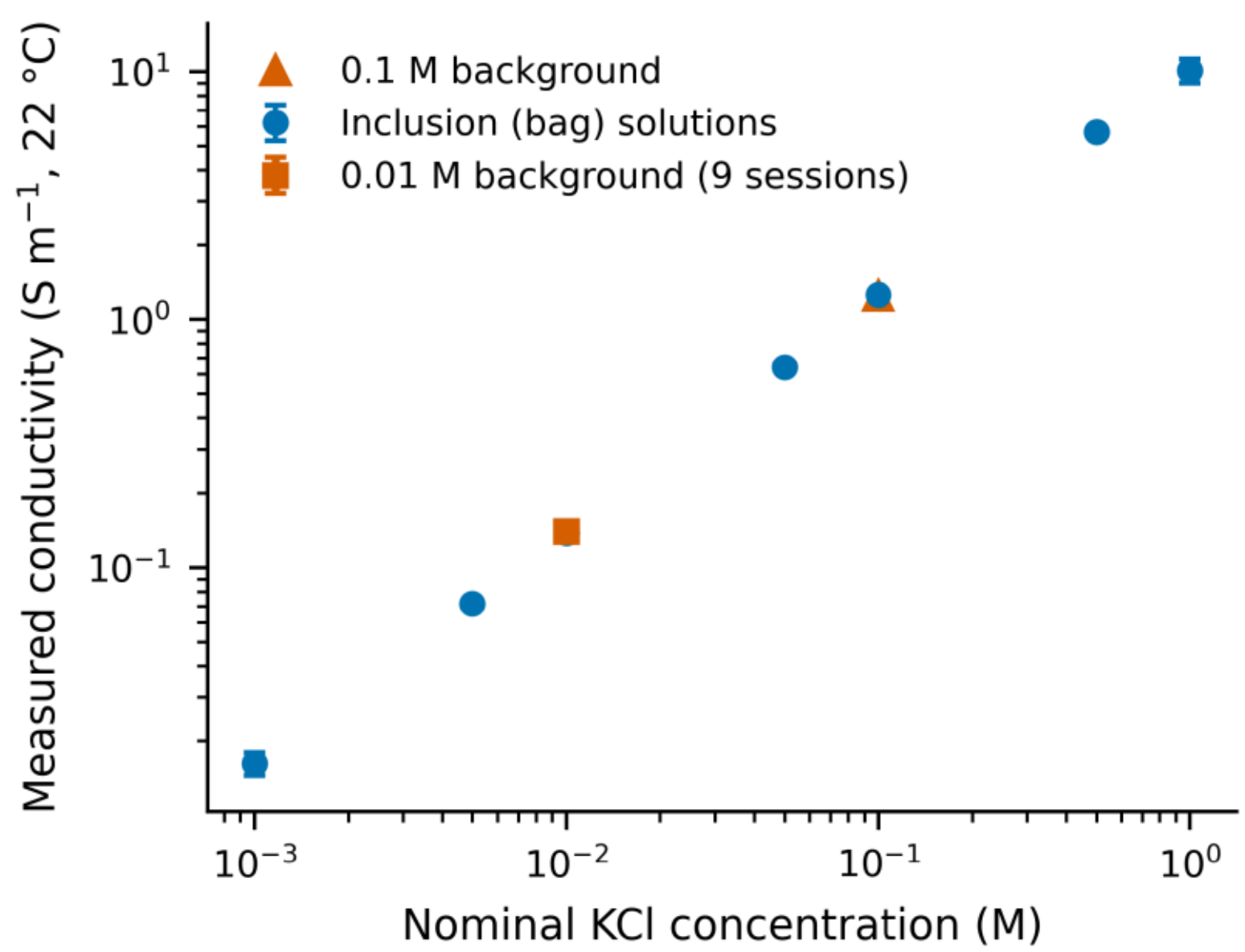

**Figure S3.** Measured conductivity (S $m^{-1}$, 22 °C) of all KCl solutions by nominal concentration; session-resolved measurements, error bars = SD across sessions.

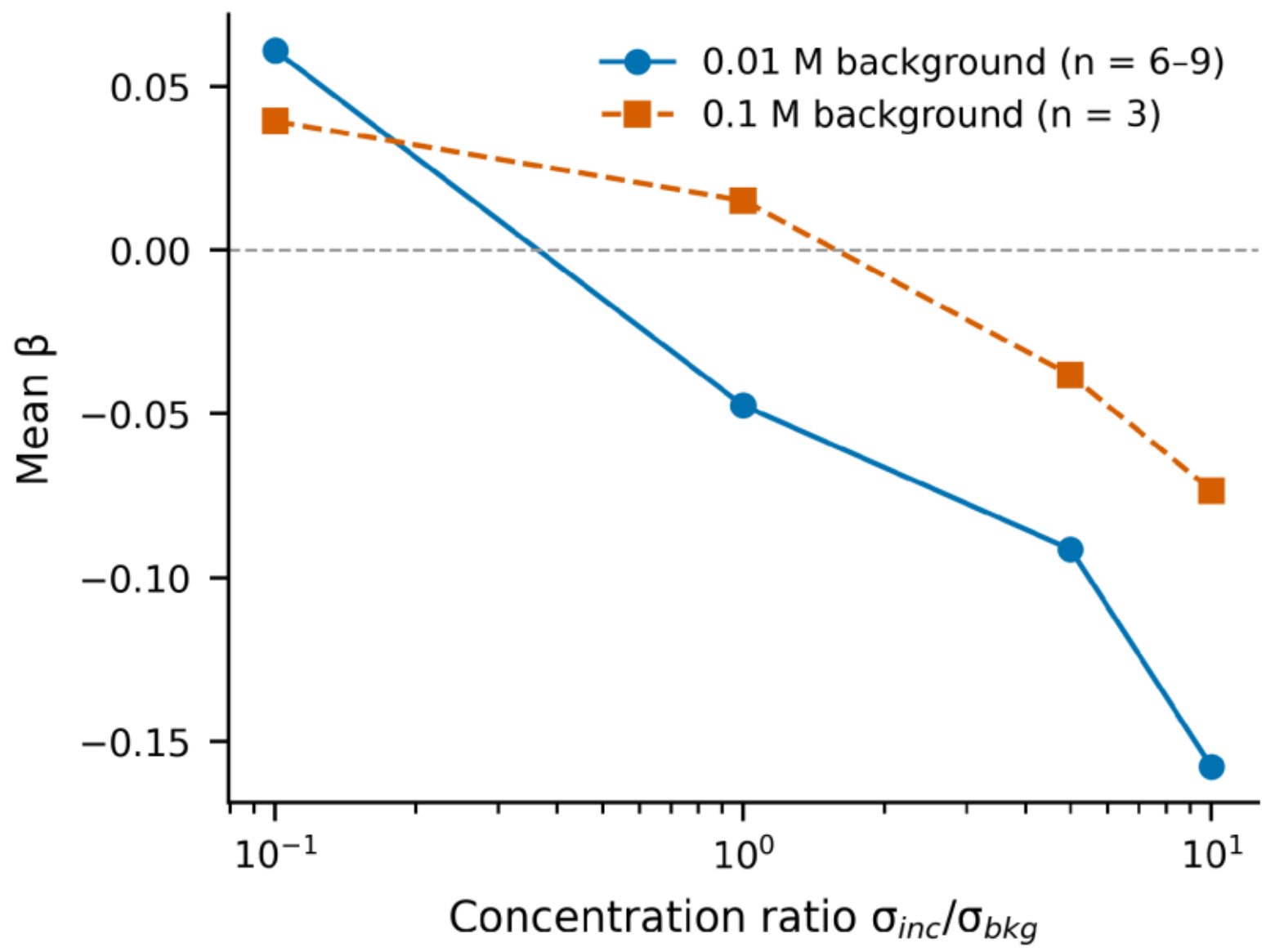

**Figure S4.** Cross-electrolyte transfer of the signed contrast projection: mean β versus concentration ratio in 0.01 M and 0.1 M KCl backgrounds; the rank ordering and zero-crossing are preserved.